\RequirePackage{rotating}
\documentclass[a4paper,fleqn,usenatbib,useAMS]{mnras}

\usepackage{graphicx}	
\usepackage{amsmath}	
\usepackage{amssymb}	
\usepackage{multicol}   
\usepackage{bm}		
\usepackage{pdflscape}	
\usepackage{caption}
\usepackage[T1]{fontenc}
\usepackage{ae,aecompl}
\usepackage{newtxtext,newtxmath}
\title[Bright \textit{Kepler} Giants]{Seismic and spectroscopic analysis of 9 bright red giants observed by \textit{Kepler}}

\author[Coelho et al.]{
H.~R.~Coelho$^{1}$\footnotemark,
A.~Miglio$^{2,3}$, 
T.~Morel$^{4}$,
N.\,Lagarde$^{5}$,        
D.~Bossini$^{6}$, 
W.~J.~Chaplin$^{7,8}$, \newauthor 
S.\,Degl'Innocenti$^{9,10}$,       
M.\,Dell'Omodarme$^{9}$,
R.~A.\,Garcia$^{11}$,                 
R.\,Handberg$^{8}$,      
S.\,Hekker$^{12,13}$, \newauthor
D.\,Huber$^{14}$,
M.\,N.\,Lund$^{8}$,
S.\,Mathur$^{15,16}$,
P.\,G.\,Prada Moroni$^{9,10}$, 
B.\,Mosser$^{17}$,                     
A.\,Serenelli$^{18,19}$,   \newauthor
M.\,Rainer$^{20}$,    
J.\,D.\,do Nascimento Jr.$^{1,21}$,         
E.\,Poretti$^{20}$,
P.\,Mathias$^{22}$,
G.\,Valle$^{9}$,   
P.\,Dal Tio$^{23,24}$,   \newauthor 
and T.\,Duarte$^{25}$
\\ $^1$ Universidade Federal do Rio Grande do Norte, UFRN, Departamento De F\'isica CP 1641, 59072-970, Natal, Brazil
\\ $^2$ Dipartimento di Fisica e Astronomia, Universit\`a degli Studi di Bologna, Via Gobetti 93/2, I-40129 Bologna, Italy
\\ $^3$ INAF - Osservatorio di Astrofisica e Scienza dello Spazio, via P. Gobetti 93/3, 40129 Bologna, Italy
\\ $^4$ Space sciences, Technologies and Astrophysics Research (STAR) Institute, Universit\'e de Li\`ege, Quartier Agora, All\'ee du 6 Ao\^ut 19c, B\^at. B5C, B4000-Li\`ege, Belgium
\\ $^5$Laboratoire d'Astrophysique de Bordeaux, Universit\'e Bordeaux, CNRS, B18N, All\'ee Geoffroy Saint-Hilaire, 33615 Pessac, France
\\ $^6$ Instituto de Astrof\'isica e Ci$\hat{e}$ncias do Espa\c{c}o, Universidade do Porto, CAUP, Rua das Estrelas, 4150-762 Porto, Portugal
\\ $^7$ School of Physics \& Astronomy, University of Birmingham, Edgbaston, Birmingham, B15 2TT, UK 
\\ $^8$ Stellar Astrophysics Centre (SAC), Department of Physics and Astronomy, Aarhus University, Ny Munkegade 120, DK-8000 Aarhus C, Denmark
\\ $^9$ Dipartimento di Fisica "Enrico Fermi", Universit\`a  di Pisa, Largo Pontecorvo 3, Pisa I-56127, Italy
\\ $^{10}$ INFN, Sezione di Pisa, Largo Pontecorvo 3, I-56127, Pisa, Italy
\\ $^{11}$ Universit\'e Paris-Saclay, Universit\'e Paris Cit\'e, CEA, CNRS, AIM,  91191, Gif-sur-Yvette, France
\\ $^{12}$ Center for Astronomy (ZAH/LSW), Heidelberg University, K\"{o}nigstuhl 12, 69117 Heidelberg, Germany
\\ $^{13}$ Heidelberg Institute for Theoretical Studies (HITS) gGmbH, Schloss-Wolfsbrunnenweg 35, 69118 Heidelberg,Germany
\\ $^{14}$ Institute for Astronomy, University of Hawai'i, 2680 Woodlawn Drive, Honolulu, HI 96822, USA
\\ $^{15}$ Instituto de Astrofi\'isica de Canarias, La Laguna, Tenerife, Spain
\\ $^{16}$ Departamento de Astrof\'isica, Universidad de La Laguna, La Laguna, Tenerife, Spain
\\ $^{17}$ LESIA, Observatoire de Paris, PSL Research University, CNRS, Universit\'e Pierre et Marie Curie, Universit\'e Paris Diderot, 92195 Meudon, France
\\ $^{18}$ Institute of Space Sciences (ICE, CSIC), Carrer de Can Magrans S/N, Cerdanyola del Valles, E-08193, Spain
\\ $^{19}$ Institut d'Estudis Espacials de Catalunya, C/Gran Capita 2-4, Barcelona, E-08034, Spain
\\ $^{20}$ Osservatorio Astronomico di Brera, via E. Bianchi 46, I-23807 Merate (LC), Italy
\\ $^{21}$ Harvard-Smithsonian Center for Astrophysics, Cambridge, MA 02138, USA
\\ $^{22}$ IRAP, Universit\'e de Toulouse, CNRS, UPS, CNES, 57 avenue d'Azereix, 65000, Tarbes, France
\\ $^{23}$ Dipartimento di Fisica e Astronomia Galileo Galilei, Universit\`a di Padova, Vicolo dell'Osservatorio 3, I-35122 Padova, Italy
\\ $^{24}$ Osservatorio Astronomico di Padova - INAF, Vicolo dell'Osservatorio 5, I-35122 Padova, Italy
\\ $^{25}$ Universidade Federal do Cariri, Juazeiro do Norte, 63048-080, Cear\'a, Brazil
}

\begin{document}

\maketitle

\begin{abstract}

Photometric time series gathered by space telescopes such as CoRoT and \textit{Kepler} allow to detect solar-like oscillations in red-giant stars and to measure their global seismic constraints, which can be used to infer global stellar properties (e.g. masses, radii, evolutionary states). Combining such precise constraints with photospheric abundances provides a means of testing mixing processes that occur inside red-giant stars. In this work, we conduct a detailed spectroscopic and seismic analysis of nine nearby (d < 200 pc) red-giant stars observed by \textit{Kepler}. Both seismic constraints and grid-based modelling approaches are used to determine precise fundamental parameters for those evolved stars. We compare distances and radii derived from Gaia Data Release 3 parallaxes with those inferred by a combination of seismic, spectroscopic and photometric constraints. We find no deviations within errorsbars, however the small sample size and the associated uncertainties are a limiting factor for such comparison. We use the period spacing of mixed modes to distinguish between ascending red-giants and red-clump stars. Based on the evolutionary status, we apply corrections to the values of $\Delta\nu$ for some stars, resulting in a slight improvement to the agreement between seismic and photometric distances. Finally, we couple constraints on detailed chemical abundances with the inferred masses, radii and evolutionary states. Our results corroborate previous studies that show that observed abundances of lithium and carbon isotopic ratio are in contrast with predictions from standard models, giving robust evidence for the occurrence of additional mixing during the red-giant phase.

\end{abstract}


 \section{Introduction}
 \label{sec:intro}
 \footnotetext{E-mail: hugo.pesquisa@gmail.com}

Asteroseismology of red-giant stars has proven to be a very successful tool to place tight constraints on fundamental stellar properties, including radius, mass, evolutionary state, internal rotation and age (see e.g. \citealt{Chaplin13}, \citealt{Hekker17}, and references therein). While ideally one would conduct an analysis of individual frequencies on a star-by-star basis, the very large number of stars observed by CoRoT \citep{Baglin06}, \textit{Kepler} \citep{Boru10}  and TESS \citep{Ricker15} makes this impractical with current analysis procedures. Most studies have so far relied on using the so-called global seismic parameters: the frequency of maximum oscillation power $\nu_{\rm{max}}$ and the average large separation $\langle \Delta\nu \rangle$. Such parameters are related to global stellar properties and can be combined with estimations of surface temperature in the so-called scaling relations to infer masses and radii for a large sample of field stars \citep[see, e.g.,][]{Miglio09, Kalli10, SilvaAguirre11}. Scaling relations can be combined with apparent magnitudes and effective temperature to derive seismic estimations for distances \citep[e.g. see][]{Miglio12A, Miglio13a, SilvaAguirre12, Rodrigues14, Mathur16}. This is particularly useful for targets for which one cannot rely on high-precision (better than a few percent) parallaxes. Additionally, \citet{Huber17} have shown that seismic distances will be more precise than end-of-mission \textit{Gaia} parallaxes beyond $\sim$ 3 kiloparsecs. It has thus become increasingly important to test the scaling relations and the seismically inferred distances by using bright giants with accurate parallaxes and robust spectroscopic estimations of surface temperature. 

The evolution of the surface chemical abundances in low mass red-giant stars is not entirely understood. Spectroscopic studies have shown a large number of cases where the observed surface abundances do not match the predictions made by stellar evolutionary models. Various extra-mixing mechanisms were proposed to explain this unexpected chemical pattern \citep[e.g see][]{Charb07, Charb10, Lagarde12, Busso07, Deni11}. One of the most important mixing episodes, that causes a noticeable change of the surface composition of low-mass stars, takes place at the so-called bump in the luminosity function on the red giant branch (RGB), which occurs when the hydrogen-burning shell encounters the sharp discontinuity in molecular weight left by the receding convective envelope. The net result is a decrease of the surface abundances of certain elements such as Lithium and Carbon, with an increase of Nitrogen \citep[e.g. see][]{Charb20, Lagarde19, Takeda19, Aguilera22, Magrini21}. 

Thermohaline mixing has been proposed to play an important role to explain the observed chemical surface pattern of low-mass RGB stars \citep[e.g.][]{Stan10, Deni10, Charb10}. Such instability is the direct effect of the inversion of mean molecular weight caused by the $^{3}\rm{He}(^{3}\rm{He},2p)^{4}\rm{He}$ reaction that occurs in the layers between the hydrogen-burning shell and the convective envelope. This mixing process is relevant both on the ascending red giant branch and during Helium-burning phase. Moreover, mixing induced by rotation has an effect on the internal chemical structure during the main sequence phase, revealed later during the ascending RGB phase after the first dredge-up, and it provides an explanation for certain abundance patterns observed at the surface of RGB stars. However, rotational-induced mixing alone does not account for enough mixing of chemicals to explain the abundance pattern of low-mass RGB stars observed around the luminosity bump \citep[e.g.][]{Pala06, Charb10, Charb20}.

The combination of seismic and spectroscopic constraints can be used to quantify the efficiency of the extra-mixing processes that occur inside red giant stars. Spectroscopy provides information about surface chemical properties and temperatures, while asteroseismology can give us information on stellar interiors and precise estimations of stellar mass, radius, and evolutionary state. A study of this kind was conducted by \cite{Lagarde15} for CoRoT stars using stellar models that incorporate the effects of rotation and thermohaline mixing. Despite the small size of the sample, different methods used to obtain estimations of stellar mass and radius show a good agreement, within standard errors. However, in most stars in the CoRoT sample, seismic constraints were not stringent enough to, e.g., constrain the evolutionary state. \textit{Kepler}'s longer duration, higher S/N observations have the potential to allow for a more effective combination of spectroscopy and asteroseismology.

In this work, we combined asteroseismology and spectroscopy to present a study of 9 nearby red giant stars observed by the \textit{Kepler} space telescope in its long-cadence mode \citep{Jenkins10}. We use this sample to test estimations of global stellar parameters obtained through the use of global seismic parameters coupled with robust spectroscopic estimations of surface temperature. The period spacing of mixed modes is extracted to better constrain the evolutionary stage of our stars. Grid based modelling is employed to obtain another set of global stellar parameters to be compared with the ones inferred from scaling relations. We test estimations of stellar distances obtained with seismic parameters by comparing them with accurate values of parallaxes obtained by \textit{Gaia} Data Release 3 (DR3). Finally, the high-quality spectroscopic data offer the opportunity to test non-canonical mixing processes inside evolved stars by taking the observed chemical abundances of surface elements and comparing them to theoretical predictions from models that account for such mixing events. 

This paper is organized as follows: In Section \ref{spec}, we discuss the methodology used in obtaining the spectroscopic parameters. In Section \ref{seismo} we discuss the seismic analysis, beginning with light-curve preparation and estimation of global seismic parameters, followed by the use of seismic scaling relations to infer global stellar properties. In Section \ref{grid}, we use grid-based pipelines to infer global stellar properties and compare them to independent determinations of radii/distances based on astrometric constraints from \textit{Gaia}. In Section \ref{discu}, we combine the spectroscopic and seismic data to test current models of internal mixing in the red-giant phase. 

\section{Spectroscopic analysis}
\label{spec}

High signal-to-noise ratio (S/N $\ge$ 200) optical spectra were necessary for our project. A Directory Discretionary Time proposal was submitted to obtain such spectra with the Narval spectrograph \citep[R$\sim$75000,][]{Auri03} mounted at the Telescope Bernard Lyot (Pic du Midi, France). Our sample is made up of 9 bright red-giant targets in the \textit{Kepler} field with high quality photometric data. The availability of accurate parallaxes, interstellar extinction and magnitudes for all targets would be crucial to obtain accurate stellar luminosities. Additionally, the sample is in the mass range where it is claimed that thermohaline instability coupled with the effects of rotation is the most efficient transport processes for chemical elements. Nearly two hours of telescope time were used to collect the requested spectra in October 2014. Data reduction was performed with the libre-esprit package \citep{Donati97}. Initial sets of effective temperatures ($T_{\rm{eff}}$), metallicities ([Fe/H]) and surface gravities ($\log g$) were obtained independently by three different teams. This initial, unconstrained analysis does not assume the seismic $\log g$ as a prior.

\begin{enumerate}

\item
Spectral synthesis software SME \citep[Spectroscopy Made Easy,][]{Val96}: Using the wavelength ranges of 5160-5190 \AA, 6000-6030 \AA, 6050-6070 \AA, 6100-6118 \AA, 6121-6140 \AA, 6142-6159 \AA\ and 6160-6180 \AA\ \citep{Val05}. The spectral regions used were normalized to the continuum and the software was run several times with different starting values in order to be sure of the stability of the results. Different stellar models (ATLAS9, ATLAS12 and MARCS) were used with SME: all the values found are usually in good agreement with each other. We eventually used the ATLAS9 model atmospheres.

\item
We also employed a spectroscopic analysis based on the equivalent width method and excitation-ionization balance of iron lines assuming one-dimensional (1D) geometry under local thermodynamic equilibrium (LTE). This method is based on the strength of a large number of selected \ion{Fe}{i} and \ion{Fe}{ii} lines. To measure the equivalent widths, we used the automated code ARES \citep{Sousa07}. The abundance of each iron line is determined using a line list retrieved from the VALD database \citep{Kupka00} and a model atmosphere. For the calculation of iron abundances, we used Kurucz atmospheric models \citep{Castelli04} with equivalent widths measurements of \ion{Fe}{i} and \ion{Fe}{ii} lines combined with the spectrum synthesis code MOOG \citep{Sneden73}. A set of stellar parameters ($T_{\rm{eff}}$, [Fe/H], $\log g$ and micro turbulence) are used as an initial guess to determine the iron abundance for each line. For first guesses, we used atmospheric parameters from the red giant star Arcturus and the Sun. This initial guess of stellar parameters are then iterated until excitation-ionization balance of iron is reached simultaneously.

\item
MOOG was also used with the Kurucz plane-parallel atmospheric models computed using ATLAS9 code ported into Linux. The methodology is identical to that carried out for CoRoT red giants by \cite{Morel14} with a number of minor improvements included in the analysis, as presented in \cite{Cam17}, to which the reader is referred for more details.

\end{enumerate}

We chose to use the temperature and metallicity values obtained through the third method because they provided surface gravities that closely matched the seismic estimations. The seismic surface gravities were obtained by combining $\nu_{\rm max}$ and $T_{\rm eff}$ in the scaling relations (see Sect. \ref{seismo} below).

\begin{figure}

 \centerline{\includegraphics[scale=0.50]{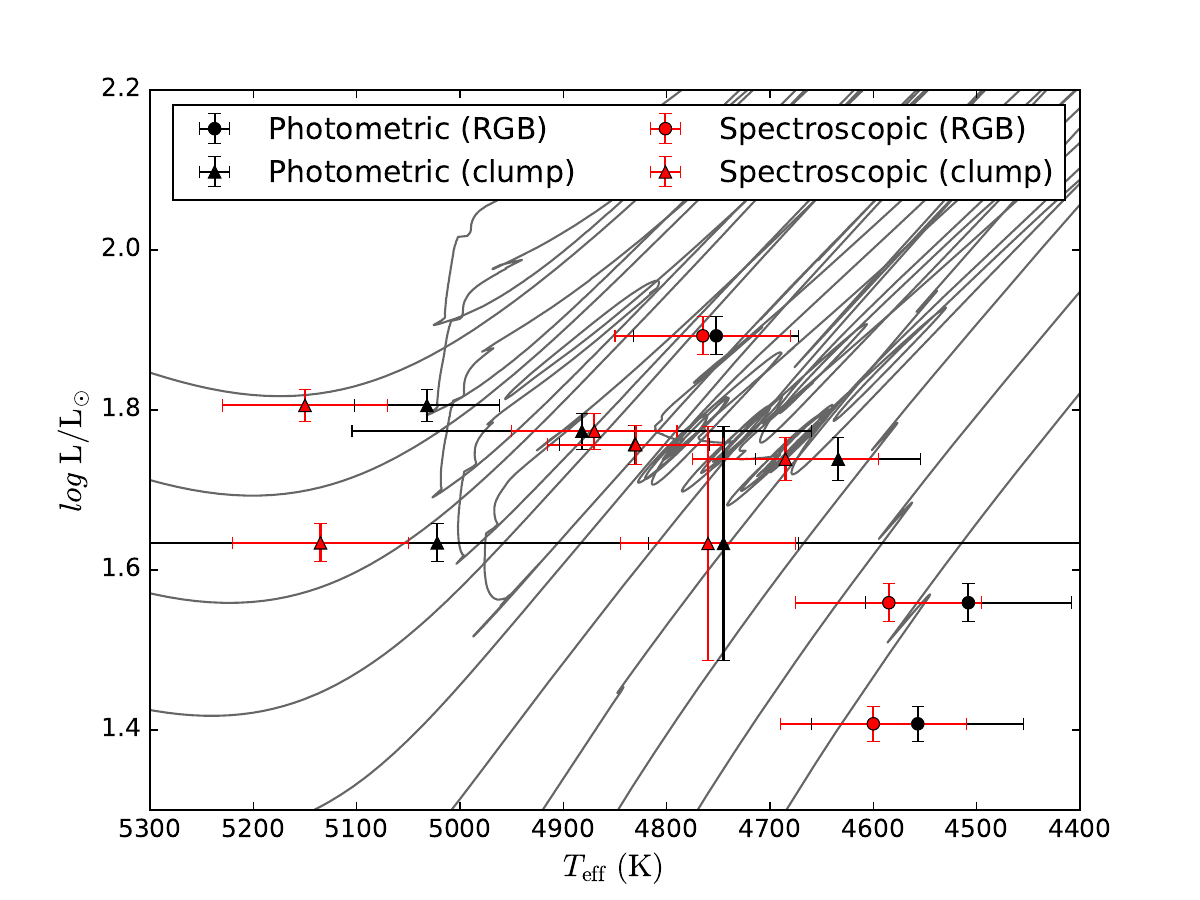}}

 \caption{HR diagram comparing the photometric temperatures (black) and the spectroscopic temperatures from the unconstrained analysis (red). Three stars in the ascending RGB phase are shown as circles, while triangles shown core Helium-burning clump stars. The luminosities are calculated using \textit{Gaia} DR3 parallaxes. One of the stars has a very high error on the B-V colour used to compute the photometric temperature. The tracks showed here are taken from the grid presented in \citet{Rodrigues17}. The range in mass covered by the tracks goes from 1.1$\rm M_{\odot}$ to 2.9$\rm M_{\odot}$, in steps of $0.2M_{\odot}$.}

 \label{HRD_photo}
\end{figure}

To investigate possible systematic uncertainties on $T_{\rm{eff}}$, we decided to compare the surface temperatures from the unconstrained spectroscopic analysis with photometric temperatures estimated with MARCS models following \citet{Casa14}. We used seismic surface gravities, metallicities, E(B-V) and observed B-V colors to obtain the desired value of photometric temperature through iteration of the parameters. Results showed that the spectroscopic temperatures for the three hottest stars in our sample are higher by roughly $\sim$ 80 K when compared to the photometric temperatures calculated using E(B-V) derived with dust maps from \citet{Green18}. The use of other dust maps \citep{Drimmel03} yields similar results. The placement of these stars in the HR diagram (at solar metallicity) and their evolutionary state as inferred from seismic constraints (see Sect. \ref{period_spacing}) suggest that the spectroscopic temperatures are slightly overestimated also when compared to the predictions from stellar models (see Figure \ref{HRD_photo}), although we do not consider this as a strong argument, given the uncertainties in predicting temperatures from stellar models \citep[e.g., see][]{Cassisi2014}.

In order to correct those effects, we conducted a new constrained determination of the spectroscopic parameters by using seismic gravity as prior. Once the surface gravity is fixed to its seismic value, it is possible to estimate effective temperatures using iron lines through two distinct methods: excitation and ionization balance.

Excitation balance consists of nulling the slope between the \ion{Fe}{i} abundances and the line lower excitation potentials. This method has been used to obtain effective temperatures for red giant stars observed by CoRoT \citep{Morel14} and \textit{Kepler} \citep{Thy12}. When this is enforced for one cool star (KIC 11918397) and one warm star (KIC 9411865) with the largest discrepancy between the spectroscopic and seismic $\log g$ ($\sim 0.2$ dex), we obtain a value of $T_{\rm{eff}}$ higher by $\sim 35$ K when compared to the unconstrained values. Since this increases the temperatures even further, we decided not to adopt this approach.

The alternative method, iron ionisation balance, requires the mean \ion{Fe}{i} and \ion{Fe}{ii} abundances to be identical. As a consequence, excitation equilibrium of the \ion{Fe}{i} lines is no longer formally fulfilled (but note that it may be within the uncertainties).

When this is done for KIC 11918397, we obtain the following results: $\Delta T_{\rm{eff}} = 60$K lower (new $T_{\rm{eff}} = 4705$ K instead of 4765 K) and $\Delta [\rm{Fe/H}]$ = 0.04 dex lower. When the same is done for KIC 9411865, we obtain: $\Delta T_{\rm{eff}} = 100$ K lower (new $T_{\rm{eff}} = 5050$ K instead of 5150 K) and $\Delta [\rm{Fe/H}]$ = 0.07 dex lower.

The temperature scale becomes cooler (which is in better agreement with the photometric $T_{\rm{eff}}$) and the metallicities are slightly lower. However, those examples are for the two stars with the largest discrepancy in $\log g$, so for the other stars the corrections will be smaller. We decided to adopt the results from iron ionisation equilibrium coupled with seismic surface gravities to obtain the effective temperatures and metallicities that will be used in grid-based modelling. A summary of the seismic and spectroscopic properties of the target stars are presented in Table \ref{seismic_table} and Table \ref{spec_table}.

\begin{figure*}
 \centerline{\includegraphics[scale=0.65]{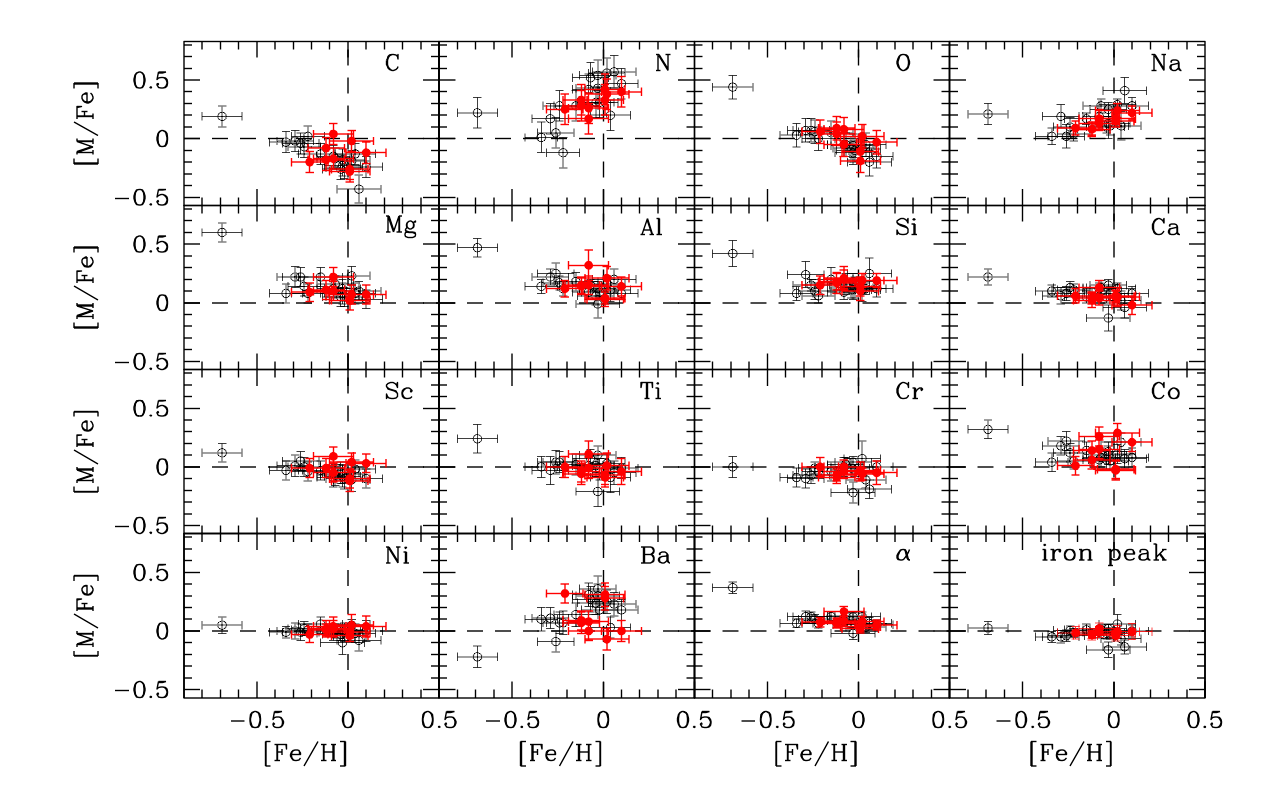}}
 \caption{Abundance ratios with respect to iron, as a function of [Fe/H]. The CoRoT red giants analysed by \citet{Morel14} are shown in black, while the stars in our sample are overplotted in red. In both cases, the results have been obtained by fixing the surface gravity to the seismic value in the spectroscopic analysis. Note, however, that excitation balance of the \ion{Fe}{i} lines was assumed in \citet{Morel14}. The mean abundance ratio of the $\alpha$-synthetised elements is defined as the unweighted mean of the Mg, Si, Ca, and Ti abundances. For the mean abundance of the iron-peak elements, we considered Cr and Ni.}
 \label{abundances}
\end{figure*}

The full set of chemical abundances can be found in Table \ref{chemical1}. Figure \ref{abundances} shows that the behaviour of the abundance ratios as a function of [Fe/H] is consistent with that found for CoRoT red giants \citep{Morel14} and, in general, with that commonly observed for thin disc dwarfs \citep[e.g.][]{Bensby14}. It is worth to mention that a robust value of Li abundance is only detected in two stars. For all other stars, the analysis provides an upper limit to the Li abundance only. 


\section{Asteroseismic data analysis}
\label{seismo}

Global stellar properties can be derived by using two key global seismic parameters:  $\nu_{\rm{max}}$ and $\Delta\nu$. The frequency of maximum oscillation power $\nu_{\rm{max}}$ is commonly assumed to scale with the atmospheric cut-off frequency, $\nu_{\rm{ac}}$ \citep{Brown91, Belka11}, which implies that, after adopting the appropriate approximations, $\nu_{\rm{ac}} \propto gT_{\rm{eff}}^{-1/2}$ \citep{Brown91, KB95}. The large separation $\langle \Delta\nu \rangle$ is defined as the average of the observed frequency spacings between consecutive overtones $\textit{n}$ with same angular degree, $\textit{l}$. The average large separation scales to a very good approximation as $\langle \rho^{1/2} \rangle$, where $\rho$ is the mean density of the star \citep[see, e.g.,][]{Tass80, Ulrich86, CD93}. Given the high quality of the photometric data, we can expect to have good estimations of $\nu_{\rm{max}}$ and $\Delta\nu$, as well as the asymptotic period spacing of gravity modes (internal gravity waves, or g modes, where buoyancy effects act as the restoring force). 

\begin{table*}
\caption{Seismic properties for the stars in our sample.}
\label{seismic_table}
\begin{tabular}{ccccc}
 \hline
 KIC & $\nu_{\rm{max}}$ & $\Delta\nu$ & $\Delta\Pi$ & Status\\
  & $\mu$Hz  & $\mu$Hz & s & \\
 \hline
1720554 & $55.73\pm 0.48$ & $5.81\pm 0.07$ & $66\pm 4$ & RGB\\
4049174 & $41.51\pm 0.29$ & $4.48\pm 0.05$ & $319.7\pm 1$ & clump\\
8752618 & $38.78\pm 0.25$ & $4.43\pm 0.06$ & $298.9\pm 1$ & clump\\
9411865 & $77.83\pm 0.41$ & $6.45\pm 0.05$ & $268.6\pm 2$ & 2nd clump$^{b}$\\
10323222 & $46.28\pm 0.35$ & $4.85\pm 0.06$ & undetected & RGB$^{a}$\\
10425397 & $32.13\pm 0.26$ & $3.92\pm 0.05$ & $324.5\pm 1$ & clump\\
11408263 & $44.23\pm 0.37$ & $4.58\pm 0.05$ & $289.4\pm 1$ & clump\\
11808639 & $97.99\pm 0.56$ & $7.93\pm 0.07$ & $226.6\pm 2$ & 2nd clump$^{b}$\\
11918397 & $32.13\pm 0.31$ & $3.51\pm 0.04$ & undetected & RGB$^{a}$\\  
\hline
\multicolumn{5}{l}{$^a$ Assumed to be RGB (see discussion in Sect. \ref{extract} for more details).}\\
\multicolumn{5}{l}{$^b$ "2nd clump" indicate stars in the secondary clump.}\\
\end{tabular}
\end{table*}

\subsection{Light curve preparation}

Light curves were constructed from pixel data \citep{Jenkins10} downloaded from the KASOC database \footnote{\url{kasoc.phys.au.dk}} using the procedure developed by S. Bloemen (private comm.) to automatically define pixel masks for aperture photometry. The extracted light curves were then corrected using the KASOC filter \citep[see][]{Handberg14}. The light curves are first corrected for jumps upon which they are concatenated. Each light curve is then median filtered using two filters of different widths, with the final filter being a weighted sum of the two filters based on the variability in the light curve. The power density spectra (PDS) used in the following seismic analysis are made from a weighted least-squares sine-wave fitting, single-sided calibrated, normalized to Parseval's theorem, and converted to power density by dividing by the integral of the spectral window \citep{Kjeld92, KF92}. 

In order to make a better version of the light curve than those provided by the Pre-search Data Conditioning (PDC) pipeline, we recreated the mask and filtered the data in another way that is more optimized for asteroseismology. The custom masks are created based on an estimate of the pixel response function for the star \citep{Bry10}, in such manner that brighter targets will have appropriately larger masks, yielding light curves that are more suited for asteroseismic analysis \citep{Handberg14}. 

\subsection{Extraction of global seismic parameters}

\begin{figure*}

 \includegraphics[width=\columnwidth]{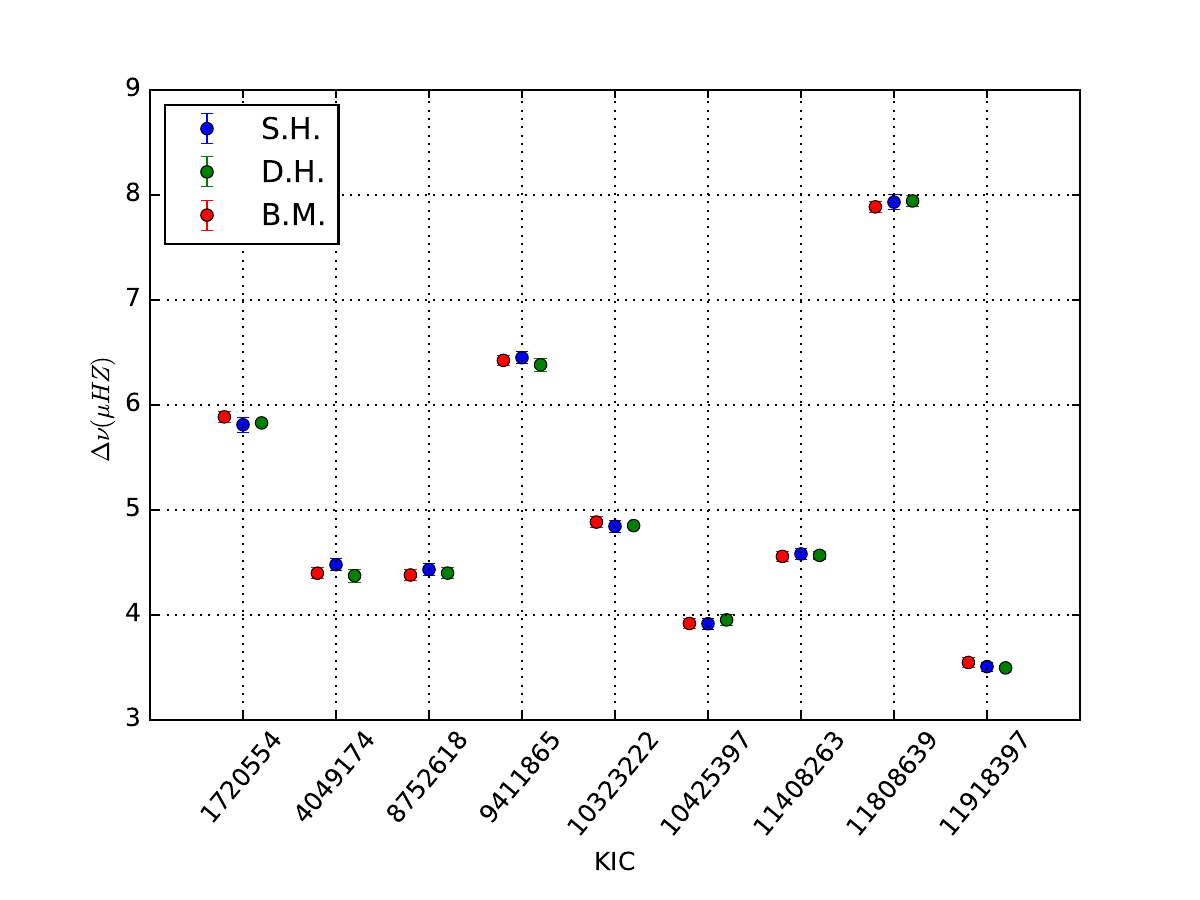}
 \includegraphics[width=\columnwidth]{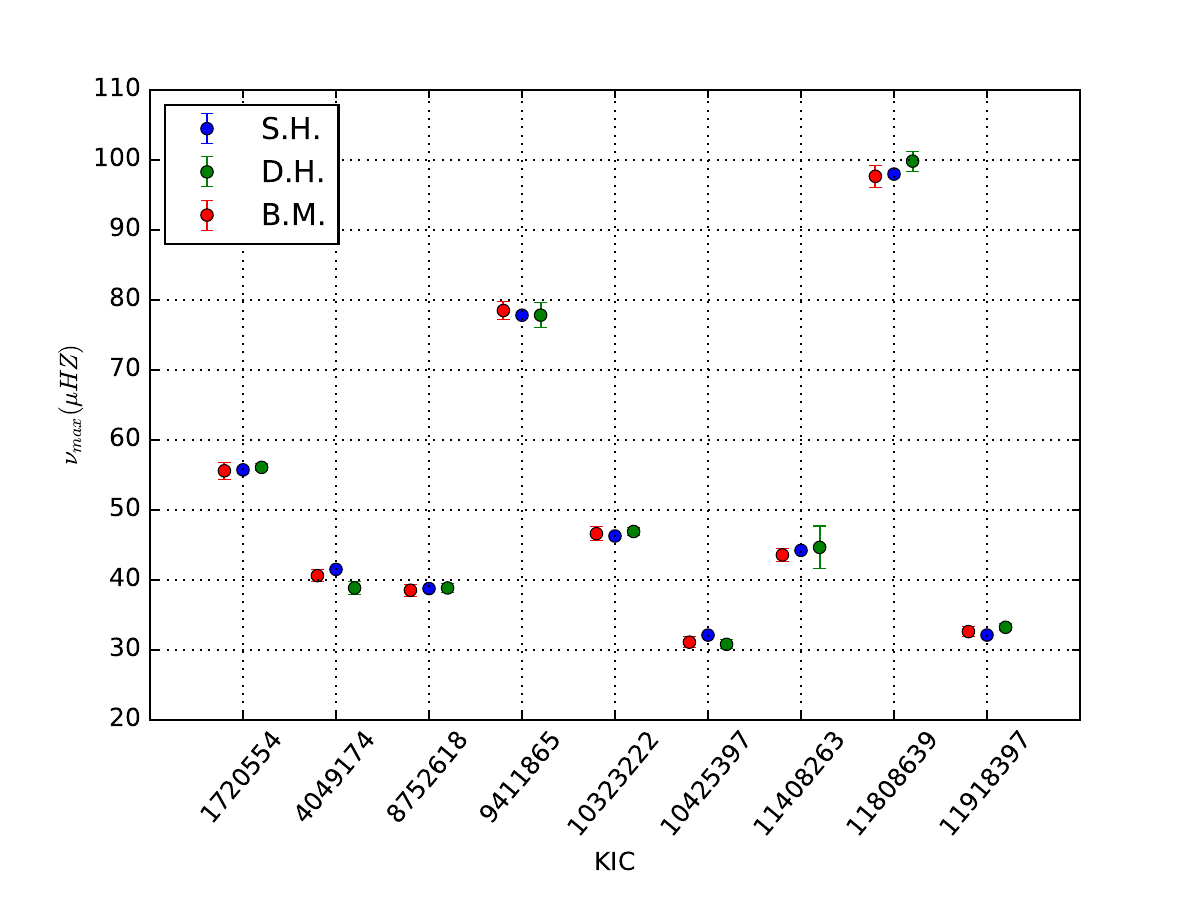}

 \caption{Left panel: Comparison between estimations of $\Delta\nu$ obtained by three different pipelines. Right panel: same as left panel, now for $\nu_{\rm max}$.}

 \label{dnu_comparison}
\end{figure*}

Three different methods have been used to detect oscillations and extract global seismic parameters $\Delta\nu$, $\nu_{\rm max}$ and the period spacing $\Delta\Pi$ from the power spectra. We refer to these methods as B.M., D.H. and S.H. from here on. 

\begin{enumerate}

\item
The methods used by B.M. are described in extensive detail in \citet{Mosser09} for the global analysis, \citet{Mosser11} for the precise $\Delta\nu$ measurement from the universal red giant oscillation pattern, \citet{Mosser12} for extraction of period spacing and \citet{Mosser14} for the determination of evolutionary status. 

\item
D.H. used the methodology described in \citet{Huber09} to extract $\Delta\nu$ and $\nu_{\rm max}$, however it should be noted that D.H. used his own detrending for the light curves, rather than the data available at the KASOC database. 

\item
S.H. was able to provide estimations of $\Delta\nu$ and $\nu_{\rm max}$ for all stars, by using the Octave pipeline (OCT) and the methodology described in \citet{Hekker10}.

\end{enumerate}

The median scatter between the seismic pipelines is $0.1\%$ in $\Delta\nu$ and $0.7\%$ in $\nu_{\rm max}$ (see Figure \ref{dnu_comparison}). The small differences between the global oscillation parameters obtained from the different asteroseismic pipelines do not have significant impact on the conclusions presented in this work. In the discussion that follows we adopted the values provided by S.H. (see Table \ref{seismic_table}), since they are closest to the median between the pipelines. 

\begin{figure}
 
 \includegraphics[width=\columnwidth]{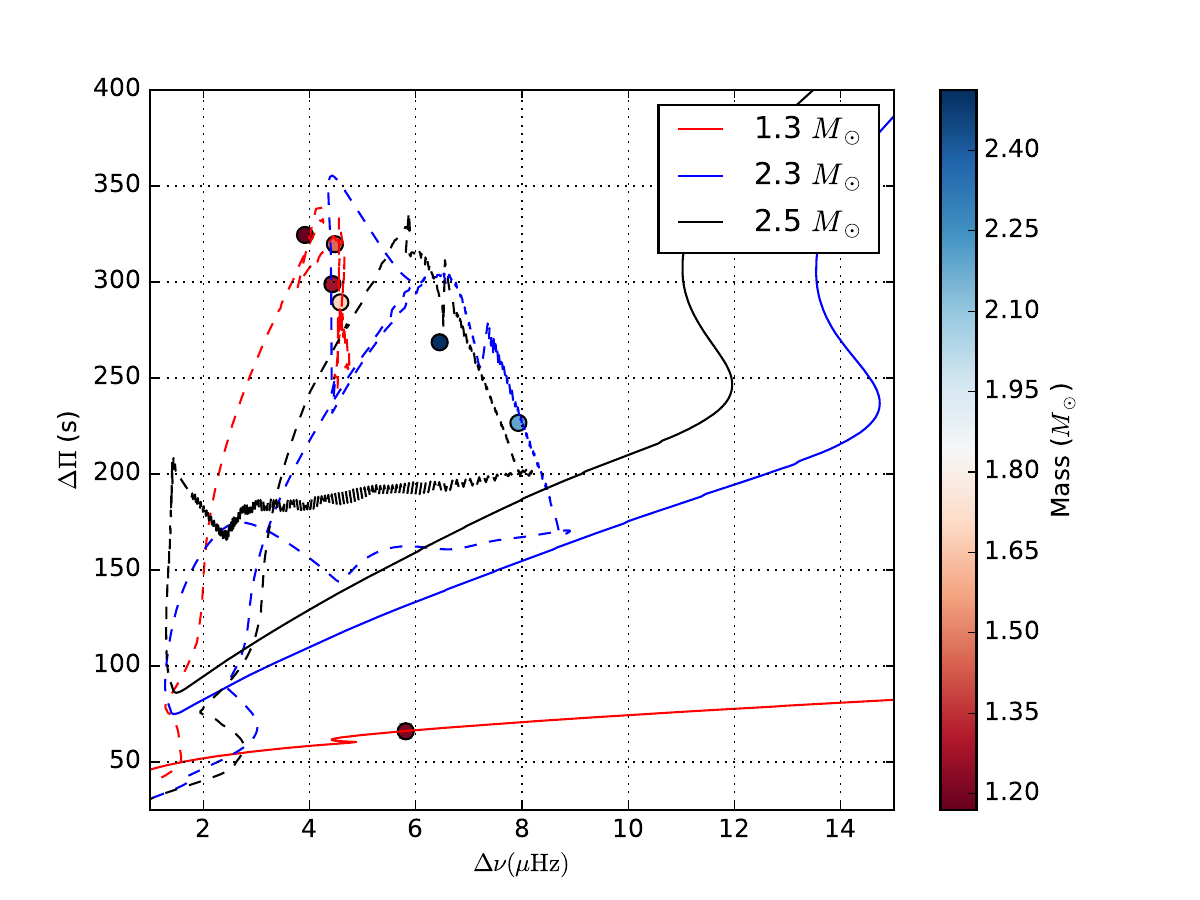}

 \caption{Period spacing ($\Delta\Pi$) versus large separation $\Delta\nu$. The coloured lines indicate evolutionary tracks computed by the Modules for Experiments in Stellar Astrophysics (MESA) code \citep{Pax11} with the convective overshooting mixing scheme described in \citet{Bos15}. The solid lines correspond to models in the ascending RGB phase, while dashed lines show models in the core Helium-burning phase. The models have solar composition and three different masses (1.30, 2.30 and 2.50 $\rm M_{\odot}$). The colour code of the dots are related to the stellar masses through the color bar on the right of the plot. Error bars in $\Delta\Pi$ and $\Delta\nu$ are smaller than the size of the points.}

 \label{period_spacing}
\end{figure}


\subsection{Asymptotic period spacing}
\label{extract}

By assuming the asymptotic approximation of high-order and low-degree modes for a non-rotating star, the period spacing of consecutive gravity modes with the same value of angular degree \textit{l} can be expressed as (e.g. see \citealt{Tass80}, \citealt{CD11}):

\begin{equation}\label{delta_pi}
\Delta\Pi_{l} = \frac{2\pi^2}{\sqrt{l(l+1)}} \bigg( \int_{g} {N\frac{dr}{r}} \bigg)^{-1} ,
\end{equation}

where $N$ is the Bruntt-V\"ais\"al\"a frequency and the integration is done on the \textit{g} mode cavity.

Gravity modes are known to have high mode inertias and, consequently, have a very low photometric amplitudes near the surface. However, as the star evolves into a red giant and the core contracts, the frequencies of gravity modes increase and eventually g modes interact with p modes (acoustic waves where gradients of pressure act as the restoring force) of the same angular degree. Such mixed modes behave as pressure modes near the surface and gravity modes near the core. 

Measuring g-mode period spacings of dipole modes allows us to clearly discriminate between red giants with similar luminosities but different evolutionary states \citep{Bed11}. Ascending red giants, characterized by a hydrogen-burning shell around an inert He-core, show typically  low values of $\Delta\Pi$ ($\backsim$ 60s) when compared to core Helium-burning clump stars ($\Delta\Pi$ $\sim$ 300s) \citep[e.g.][]{Mosser14}. 

Asymptotic period spacings have been inferred from the precise fit of the mixed mode patterns with an asymptotic expansion (\citealt{Mosser12}, \citealt{Vrard16}). Six stars are confirmed as clump stars and one star is ascending the red giant branch. The period spacings of our stars are presented in Figure \ref{period_spacing}. The observed values and inferred evolutionary status are reported in Table \ref{seismic_table}. For two stars, namelly KIC 10323222 and KIC 11918397, the period spacings are too low to be measured properly. This indicates that these stars are non-RC stars and hence, given their  $\nu_{\rm max}$, most likely RGB stars. We expect that, if these two stars were clump stars, a clear mixed mode pattern would be detectable (e.g. see \citealt{Gro14}, \citealt{Vrard16}).

\section{Inference on stellar properties}
\label{grid}

\subsection{Stellar properties using seismic scaling relations}

\begin{figure}

 \includegraphics[width=\columnwidth]{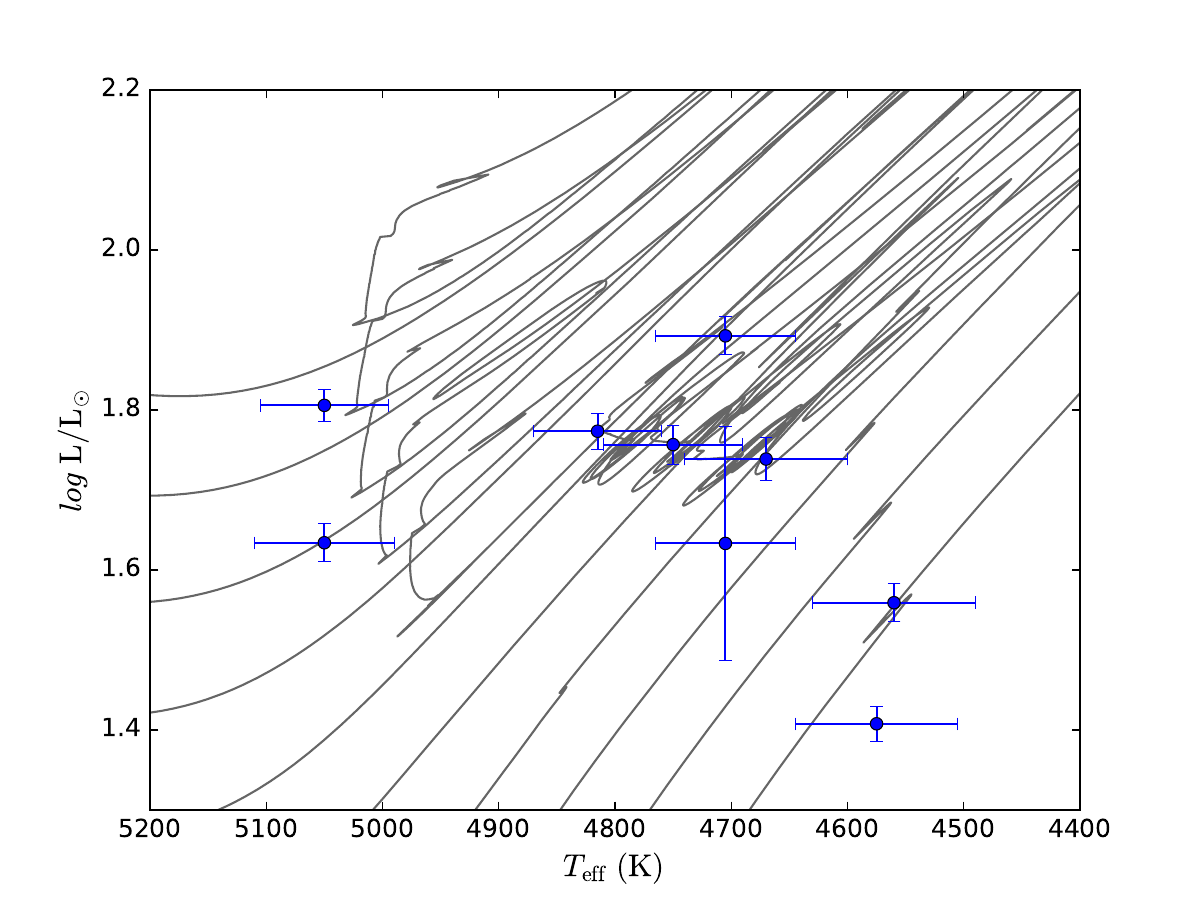}

 \caption{HR diagram for the stars in our sample. Luminosities are calculated with 2MASS K-band magnitude and parallaxes from \textit{Gaia} DR3. Effective temperatures are retrieved from the constrained spectroscopic analysis. The tracks showed here are taken from the grid presented in \citet{Rodrigues17}. The range in mass covered by the tracks goes from 1.1 $\rm M_{\odot}$ to 2.9 $\rm M_{\odot}$, in steps of $0.2 M_{\odot}$.}

 \label{HRD}
\end{figure}

To obtain photometric luminosities, we used 2MASS Ks-band magnitude with bolometric corrections calculated by using $T_{\rm{eff}}$, $\log g$ and [Fe/H] as input to the code presented in \citet{Casa14}. We accounted for the effects of interstellar extinction and reddening by using the 3D dust maps presented in \citet{Green18}. More detail about the effects of interstellar extinction and the choice made for the dust map and photometric band used in this work can be found in the appendix \ref{appendix_dust}. The bolometric corrections are used together with parallaxes from \textit{Gaia} DR3 (\citealt{Gaia23}, \citealt{Babusiaux22}) to derive luminosities (Figure \ref{HRD}). In our analysis, we applied the zero-point correction and systematic effects in parallaxes \citep{Lindegren20}. The set of spectroscopic and other non-seismic properties can be found in Table \ref{spec_table}, while Table \ref{derived_table} contains photometric data.

\begin{table*}
\caption{Spectroscopic and astrometric data for the stars in our sample. Stellar masses and $\log g$ are obtained through the use of scaling relations.}
\label{spec_table}
\begin{tabular}{cccccc}
 \hline
 KIC & $M/M_{\odot}$ & $T_{\rm{eff}}$& [Fe/H] & $\log g$ & $\varpi (Gaia)$\\\
  &  & K & dex & cgs; dex & mas  \\
 \hline
 1720554 & $1.21\pm 0.07$ & $4575\pm 70$ & $-0.08\pm 0.11$ & $2.64\pm 0.01$ & $9.506 \pm 0.067$  \\
 4049174 & $1.46\pm 0.09$ & $4670\pm 70$ & $+0.10\pm 0.11$ & $2.54\pm 0.01$ & $6.904 \pm 0.067$  \\
 8752618 & $1.27\pm 0.07$ & $4750\pm 60$ & $-0.08\pm 0.11$ & $2.49\pm 0.01$ & $6.258 \pm 0.086$  \\
 9411865 & $2.51\pm 0.10$ & $5050\pm 55$ & $+0.01\pm 0.11$ & $2.79\pm 0.01$ & $6.946 \pm 0.069$ \\
10323222 & $1.42\pm 0.08$ & $4560\pm 70$ & $+0.02\pm 0.12$ & $2.57\pm 0.01$ & $7.373 \pm 0.066$  \\
10425397 & $1.17\pm 0.07$ & $4705\pm 60$ & $-0.12\pm 0.10$ & $2.41\pm 0.01$ & $6.481 \pm 0.064$  \\
11408263 & $1.69\pm 0.09$ & $4815\pm 55$ & $-0.12\pm 0.10$ & $2.56\pm 0.01$ & $6.899 \pm 0.091$  \\
11808639 & $2.19\pm 0.09$ & $5050\pm 60$ & $+0.01\pm 0.10$ & $2.91\pm 0.01$ & $8.108 \pm  0.067$  \\
11918397 & $1.82\pm 0.11$ & $4705\pm 60$ & $-0.21\pm 0.10$ & $2.41\pm 0.01$ & $6.177 \pm 0.068$  \\
\hline
\end{tabular}
\end{table*}

To estimate stellar luminosities based on seismology, we first need to derive global stellar properties using seismic constraints. It is possible to combine the information carried by $\Delta\nu$ and $\nu_{\rm{max}}$ to obtain seismic estimates of mass and radius:

\begin{equation}\label{seismicmass}
\frac{M}{M_\odot} \simeq \left( \frac{\nu_{\rm{max}}}{\nu_{\rm{max,\odot}}} \right)^{3} \left( \frac{\Delta\nu}{\Delta\nu_{\odot}} \right)^{-4} \left( \frac{T_{\rm{eff}}}{T_{\rm{eff, \odot}}} \right)^{3/2} ,
\end{equation}
\begin{equation}\label{seismicradii}
\frac{R}{R_\odot} \simeq \left( \frac{\nu_{\rm{max}}}{\nu_{\rm{max,\odot}}} \right) \left( \frac{\Delta\nu}{\Delta\nu_{\odot}} \right)^{-2} \left( \frac{T_{\rm{eff}}}{T_{\rm{eff, \odot}}} \right)^{1/2}.
\end{equation}

\begin{figure}

 \includegraphics[width=\columnwidth]{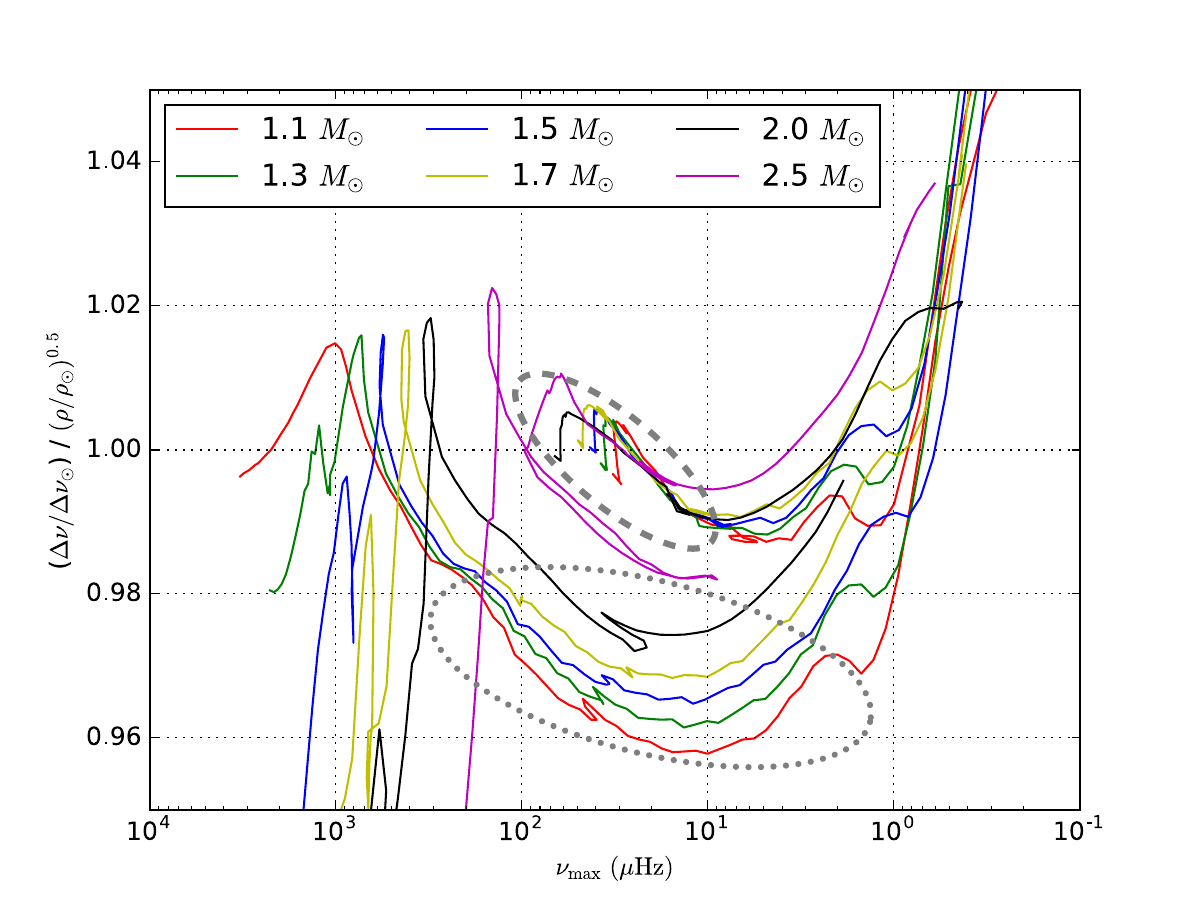}

 \caption{Ratio $\Delta\nu/\Delta\nu_{\odot}$ to $(\rho/\rho_{\odot})^{1/2}$ as a function of $\nu_{\rm{max}}$ for models computed using MESA with nearly solar metallicity ($Z = 0.01756$, $Y = 0.26556$). $\Delta\nu$ was calculated using model frequencies around $\nu_{\rm{max}}$ \citep[see][]{Rodrigues17}. The effects in low-mass clump stars (grey dashed elipse) are less than $\backsim$ 1$\%$, but effects on stars around the RGB bump (grey dotted elipse) are $\backsim$ 3$\%$.}

 \label{Dnu_scaling}
\end{figure}


In this work, we adopted solar values of $\Delta\nu_{\odot} = 135.1$ $\mu$Hz, $\nu_{\rm{max \odot}} = 3090$ $\mu$Hz and $T_{\rm{eff},\odot} = 5777$ K \citep[e.g., see][]{Chap14}. 

This set of equations has been widely used to obtain "direct" estimations of mass and radius for a given set of surface temperatures $T_{\rm{eff}}$. The limitations of the scaling relations are not fully understood yet \citep{Hekker20}, but tests have shown that, for solar-type and RGB stars, stellar radii can be as accurate as 1$\%$ to 2$\%$, while estimations of mass can be as good as 5$\%$ \citep[e.g. see][]{Guggenberger16, Rodrigues17, LiY21, LiT22, Wang23}.

Corrections to a simple scaling of $\Delta\nu \propto (\rho/\rho_{\odot})^{1/2}$ have now been proposed in the literature \citep[see ][]{White11, Miglio12B, Miglio13, Miglio16, Brogaard16, Guggenberger17, Sharma16, Theme18}. In Figure \ref{Dnu_scaling} we show the ratio $\Delta\nu/\Delta\nu_{\odot}$ to $(\rho/\rho_{\odot})^{1/2}$ as a function of $\nu_{\rm{max}}$ for models with nearly solar metallicity and a range of masses that goes from 1.1 to 2.5 $\rm M_{\odot}$. The average large separation was computed from model frequencies by using radial modes around $\nu_{\rm{max}}$, while the model frequencies take into account corrections to mitigate the impact of the so-called surface effects, since they have a significant effect on the average value of the large frequency separation (see \citealt{Rodrigues17} for a detailed description). As can be seen on the plot, the ratio deviates from unity (see e.g. \citealt{Belka13} and \citealt{Rodrigues17} for a detailed discussion). Based on this plot, we inferred corrections to our values of $\Delta\nu$ by first using period spacing ($\Delta\Pi$, see Table \ref{seismic_table}) as an indicator of evolutionary stage. For solar metalicity, stars in the clump region (located around the grey dashed elipse in Figure \ref{Dnu_scaling}) will not show large discrepancies, while stars around the RGB bump (roughly located around the grey dotted elipse in Figure \ref{Dnu_scaling}) will show larger deviations. In our sample, corrections were applied to three RGB stars, while the other six clump stars have a combination of mass and evolutionary stage that makes a correction of $\Delta\nu$ to be negligible. We note that the values of $\Delta\nu$ presented in Table \ref{seismic_table} do not take any correction into account. The adopted corrections of $\Delta\nu$ are: an increase of $3.5\%$ for KIC 1720554 ($T_{\rm{eff}}$ = 4575 K, $M \sim$ 1.2 $\rm{M_{\odot}}$) and KIC 10323222 ($T_{\rm{eff}}$ = 4560 K, $M \sim$ 1.4 $\rm{M_{\odot}}$), and an increase of $2\%$ for KIC 11918397 ($T_{\rm{eff}}$ = 4705 K, $M \sim$ 1.8 $\rm{M_{\odot}}$). At least one of the grid-based modelling codes used later will do this step self consistently, and the corrections presented here can be considered as a first iteration \citep[PARAM, see][for more details]{Rodrigues17}.

\begin{figure}
 
 \includegraphics[width=\columnwidth]{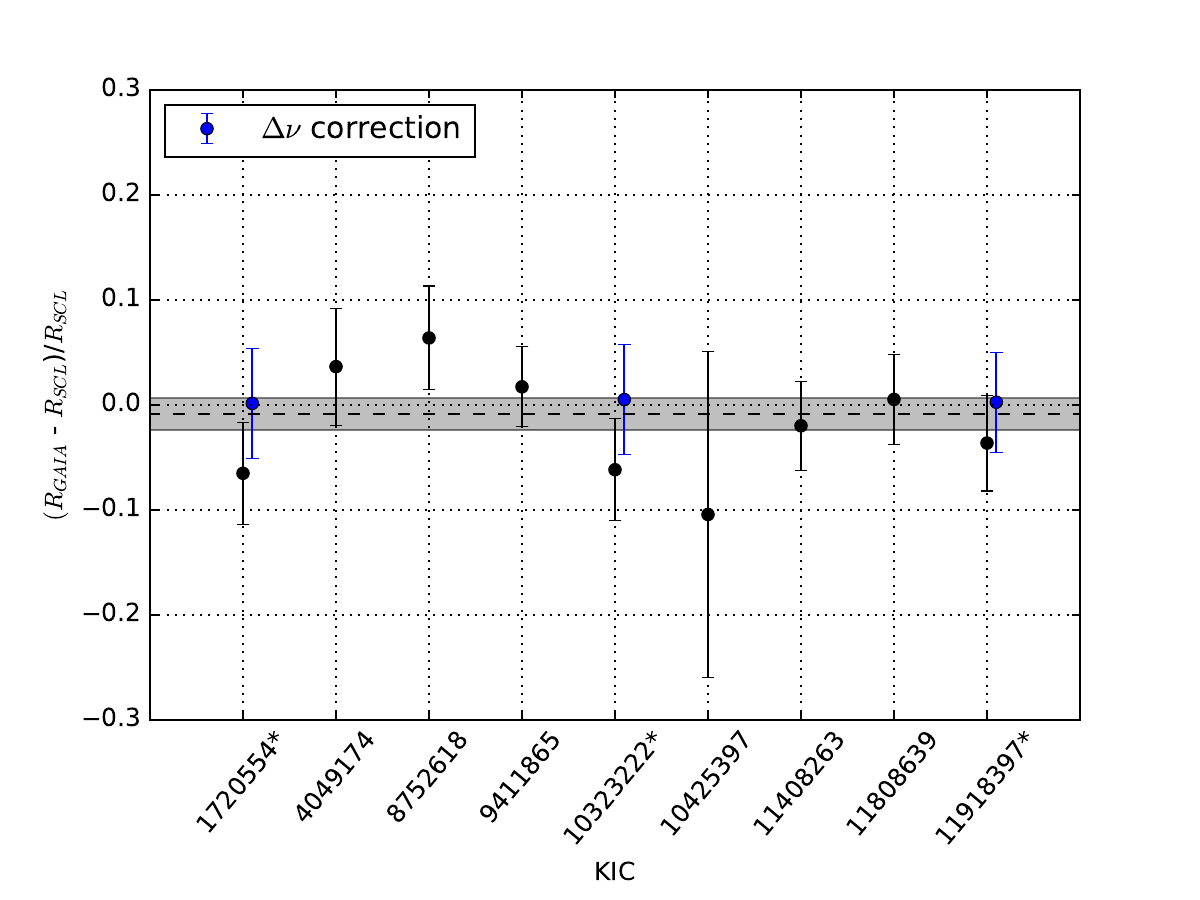} 

 \caption{Relative differences between radii derived from \textit{Gaia} DR3 parallaxes and radii derived from scaling relations (eq.  \ref{seismicradii}) for the stars in our sample. Blue dots show the three RGB stars with corrections applied to $\Delta\nu$, also indicated by asterisks in the KIC numbers. The black dashed line is the weighted average difference (without corrections in $\Delta\nu$) of $-0.0087 \pm 0.0149$.} 

 \label{res_Rscl}
\end{figure}

For the nine stars in our sample, we compared the radii calculated using magnitudes and \textit{Gaia} DR3 parallaxes with radii derived from the asteroseismic scaling relations (see equation \ref{seismicradii}), and the residual difference between the two quantities is displayed in Figure \ref{res_Rscl}. We computed the weighted average of the relative differences, and the associated statistical uncertainty, by using a Student t-distribution with a level of confidence of $68\%$ and N-1 degrees of freedom, where N is number of points. This approach is similar to the one presented in \cite{Miglio12B} and \cite{Chap98}. This was done to take into account the small number of points in the sample. The weighted average difference is $-0.0087 \pm 0.0149$, increasing to $0.0109 \pm 0.0097$ when the $\Delta\nu$ correction is applied to three RGB stars. We note that, given the small sample and the size of uncertainties, those results are statistically consistent with zero. A larger sample of stars would be ideal to conduct more robust comparison. The radii estimated by parallaxes agree reasonably well with radii derived from scaling relations for six stars in our sample, while two other stars (KIC 4049174 and KIC 8752618) have radii obtained through parallaxes which are $\sim 1-2\sigma$ higher than the seismic ones. Also note that KIC 10425397 has a much larger errorbar in its estimation of radii due to the large uncertainty in its 2MASS Ks magnitude.

We can combine magnitudes with parallaxes from \textit{Gaia} DR3 to estimate luminosities that, in turn, can be combined with spectroscopic temperatures in order to obtain stellar radii. If the radius is known, we can obtain two additional estimations of mass by combining seismic and non-seismic constraints:

\begin{equation}\label{mass1}
\frac{M}{M_\odot} \simeq \left( \frac{\Delta\nu}{\Delta\nu_{\odot}} \right)^{2} \left( \frac{R}{R_\odot} \right)^{3} ,
\end{equation}
\begin{equation}\label{mass2}
\frac{M}{M_\odot} \simeq \left( \frac{\nu_{\rm{max}}}{\nu_{\rm{max,\odot}}} \right) \left( \frac{R}{R_\odot} \right)^{2} \left( \frac{T_{\rm{eff}}}{T_{\rm{eff, \odot}}} \right)^{1/2}. 
\end{equation}

\begin{figure*}

 \includegraphics[width=\columnwidth]{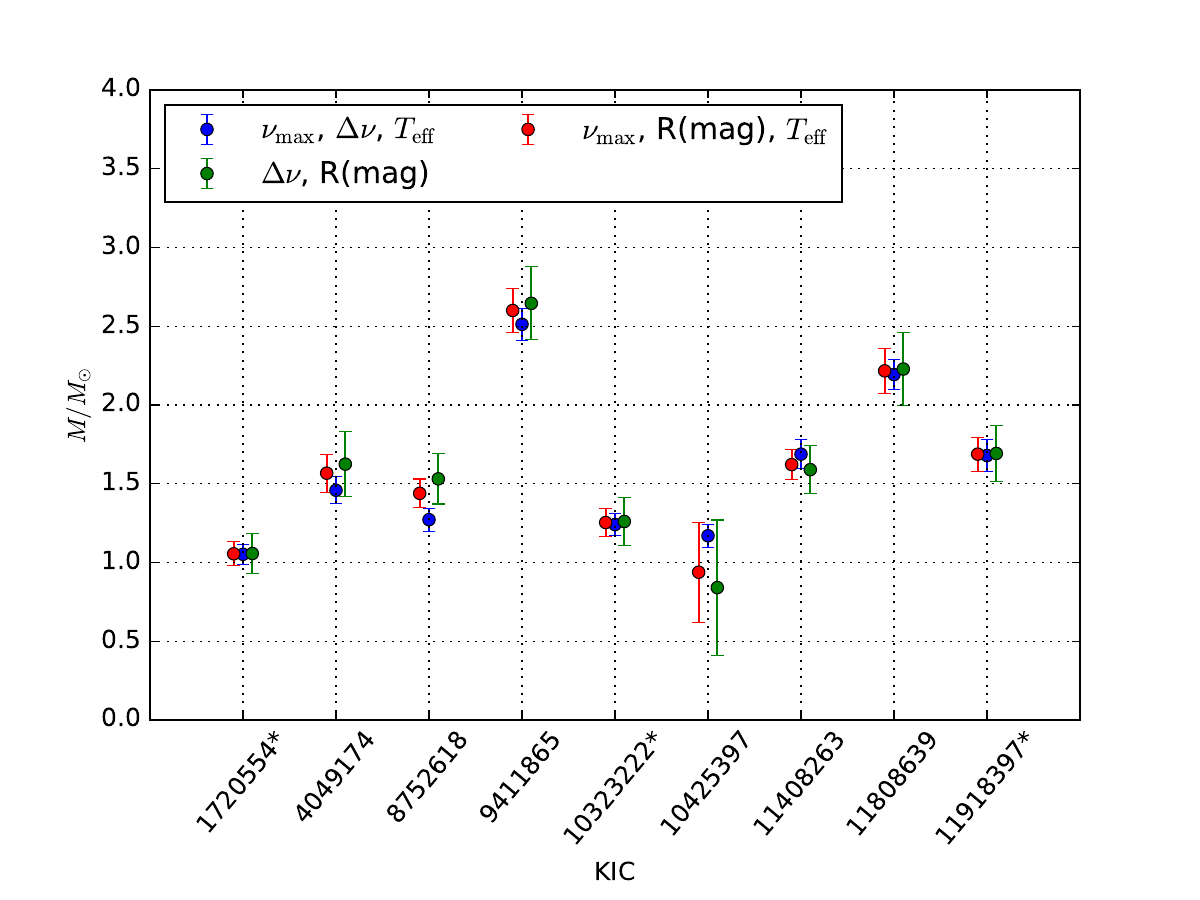}
 \includegraphics[width=\columnwidth]{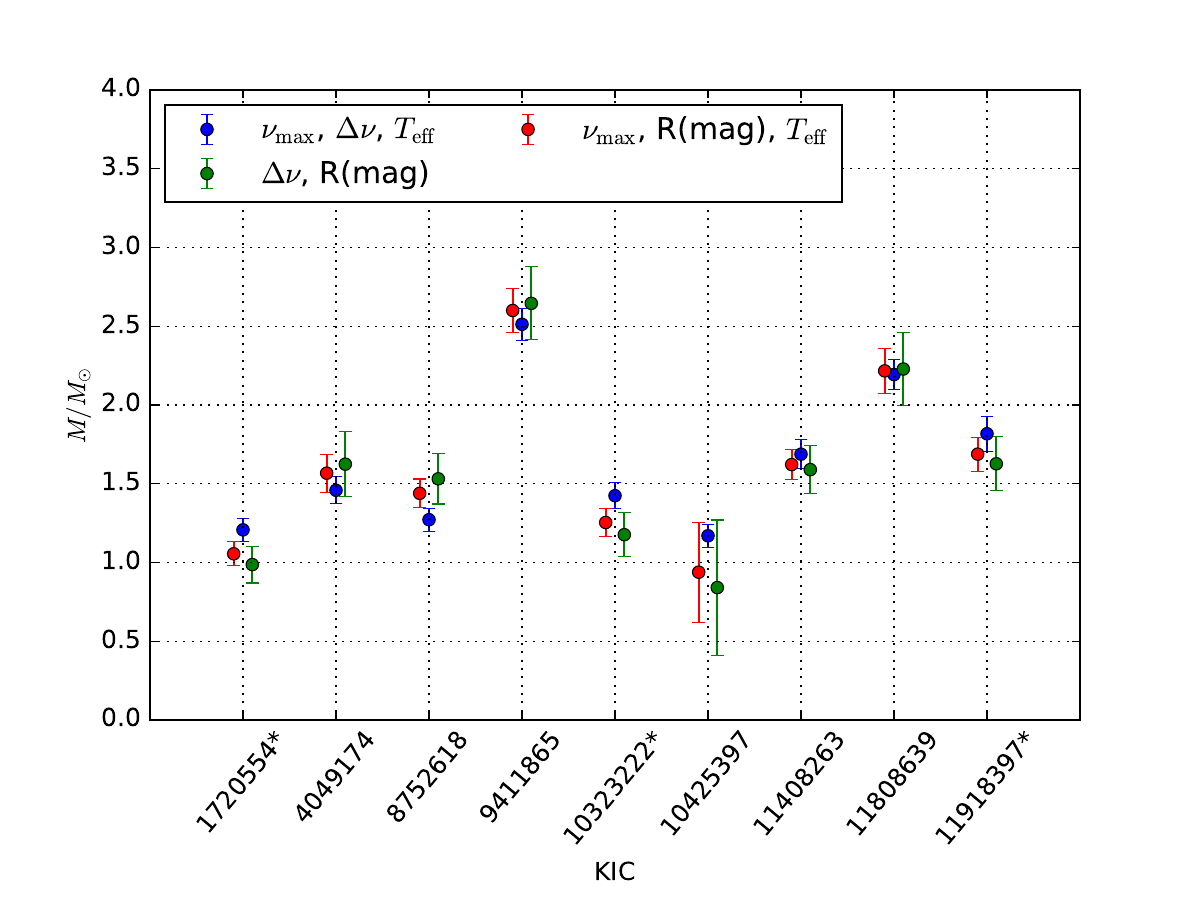}

 \caption{Left panel: Masses computed using different combinations of seismic and classical parameters. The blue dots indicate calculations made using $\nu_{\rm{max}}$, $\Delta\nu$ and the spectroscopic temperatures ($T_{\rm{eff}}$, see Equation \ref{seismicmass}). Green dots are calculated using $\Delta\nu$ and the radius obtained from \textit{Gaia} DR3 parallaxes (Equation \ref{mass1}). Red dots are calculations made using $\nu_{\rm{max}}$, spectroscopic temperatures and radius derived from \textit{Gaia} DR3 parallaxes (Equation \ref{mass2}). Right panel: Same as the left panel, now without corrections to $\Delta\nu$ applied to three RGB stars, indicated by asterisks in the KIC numbers.}

 \label{masses}
\end{figure*}

In Figure \ref{masses} we compare masses obtained using different combinations of seismic and non-seismic constraints. The typical uncertainties on mass determinations are the following: $\backsim$ 8$\%$ when using $\nu_{\rm{max}}, \Delta\nu$ and spectroscopic $T_{\rm{eff}}$ (blue dots), $\backsim$ 23$\%$ when using $\Delta\nu$ and the radius obtained from \textit{Gaia} DR3 parallaxes (green dots) and, finally, $\backsim$ 14$\%$ when using the combination of $\nu_{\rm{max}}$, spectroscopic $T_{\rm{eff}}$ and the radii derived from \textit{Gaia} DR3 parallaxes (red dots).

We then calculated a set of distances using asteroseismic parameters and compared them with distances estimated using parallaxes from \textit{Gaia} DR3. The seismic scaling relations can be combined with the Stefan-Boltzmann law, $L = 4 \pi R^{2} \sigma T_{\rm{eff}}^{4}$ (where $\sigma$ is the Stefan-Boltzmann constant), to obtain seismic distances \citep{Miglio13a}:

\begin{equation}\label{seismodist}
\log d \simeq 1 + 2.5 \log \frac{T_{\rm{eff}}}{T_{\rm{eff},\odot}} + \log \frac{\nu_{\rm{max}}}{\nu_{\rm{max,\odot}}} \\ \\
- 2\log \frac{\Delta\nu}{\Delta\nu_{\odot}} + 0.2(m_{\rm{bol}} - M_{\rm{bol},\odot}) ,
\end{equation}

\begin{figure}

 \includegraphics[width=\columnwidth]{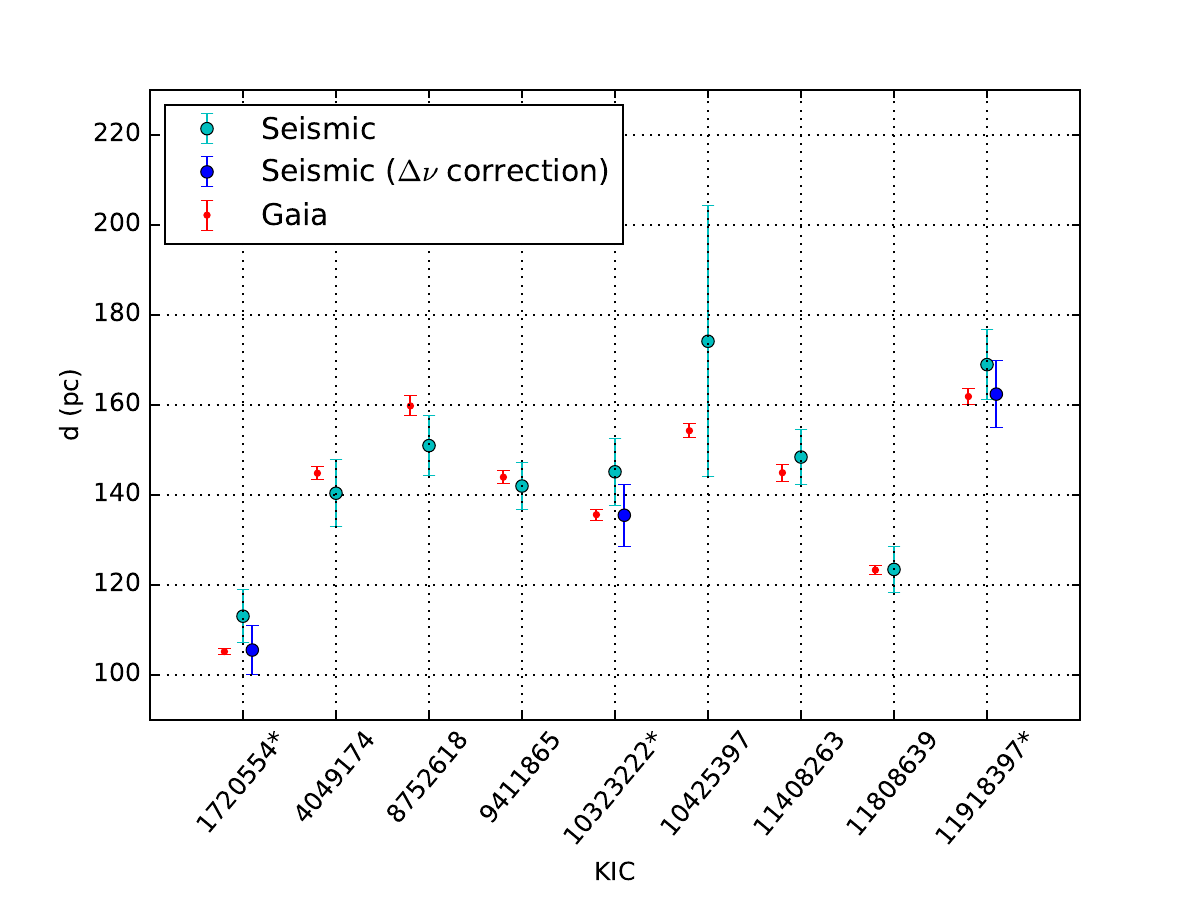}

 \caption{Comparison between the distances determined from parallaxes and distances calculated using asteroseismology. The small red dots are \textit{Gaia} DR3 distances, and light blue circles are distances determined by combining asteroseismic constraints ($\Delta\nu$, $\nu_{\rm{max}}$) with spectroscopic $T_{\rm{eff}}$. The dark blue circles are the three RGB stars with applied corrections on $\Delta\nu$, also indicated by asterisks in the KIC numbers.}

 \label{distance}
\end{figure}


\begin{figure}
 
 \includegraphics[width=\columnwidth]{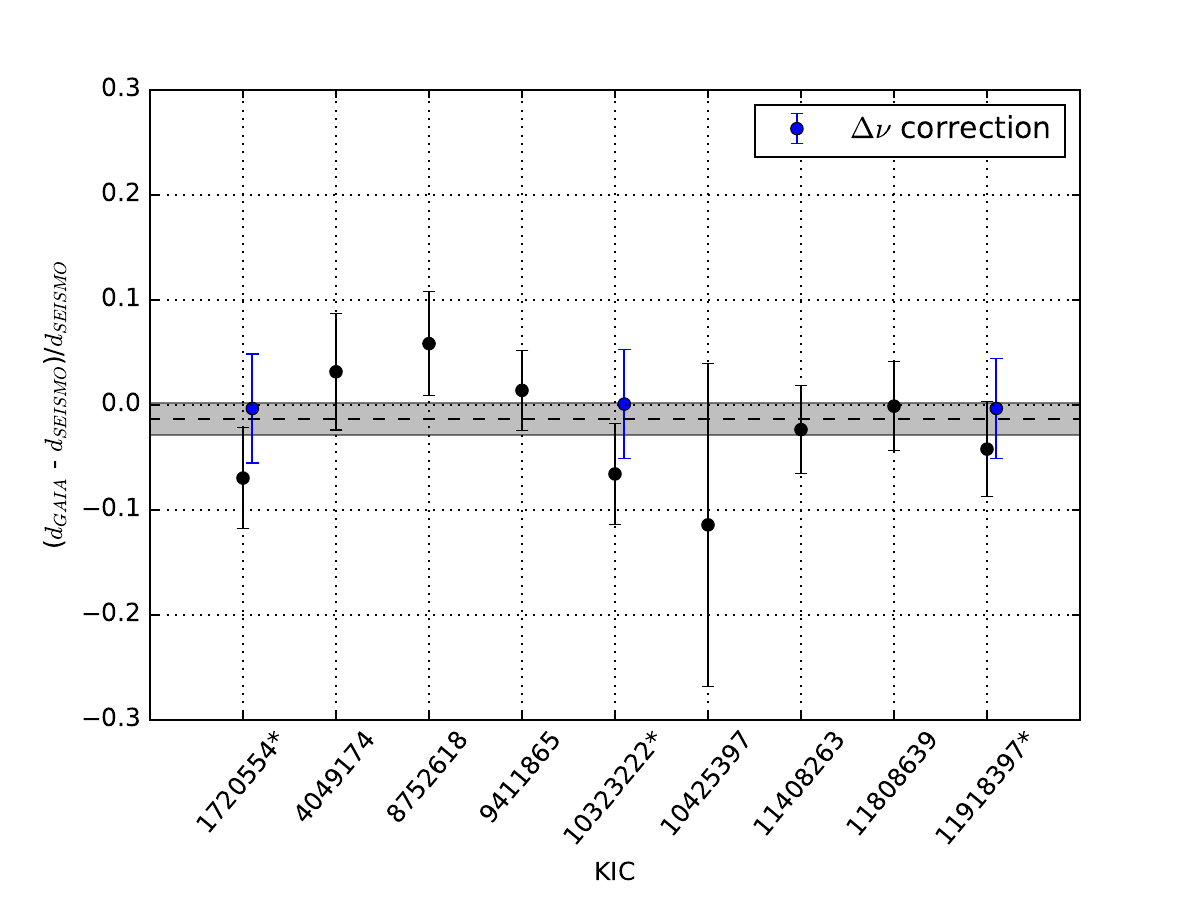} 

 \caption{Relative differences between \textit{Gaia} DR3 and seismic distances (black dots) estimated using equation \ref{seismodist}. Blue dots show the three RGB stars with corrections applied to $\Delta\nu$, also indicated by asterisks in the KIC numbers. The black dashed line is the weighted average difference (without corrections in $\Delta\nu$) of $-0.0134 \pm 0.0150$.} 

 \label{distancer}
\end{figure}


where \textit{d} is expressed in parsecs, $m_{\rm{bol}}$ is the de-reddened apparent bolometric magnitude and $M_{\rm{bol},\odot}$ is the absolute solar bolometric magnitude. Our results are shown in Figure \ref{distance}. 

We note, however, that distances derived from equation \ref{seismodist} are dependent on the estimation of extinction corrections obtained from reddening map, which may contain systematic uncertainties. Model atmospheres used to estimate bolometric corrections may also introduce systematic errors in the final computation of seismic distances. Finally, the choice of a particular $T_{\rm{eff}}$ scale may induce systematic offsets, since seismic distances scale as $T_{\rm{eff}}^{2.5}$. Therefore, our estimation of seismic distances is not completely stringent.

We have found that, for the red giant stars in our sample, the distances obtained using scaling relations are overestimated, on average, by a small amount when compared to distances derived from \textit{Gaia} DR3 parallaxes. This result is very similar to the analysis of radii mentioned above (see Figure \ref{res_Rscl}). It is important to point again that the limited amount of stars in our analysis and the size of the uncertainties are such that the difference between the two distances is statistically consistent with zero within errorbars. The relative difference between \textit{Gaia} DR3 and seismic estimations of distances are shown in Figure \ref{distancer}. The weighted average has a value of $-0.0134 \pm 0.0150$, and if the correction to $\Delta\nu$ is applied, the weighted average becomes $0.0061 \pm 0.0099$. Repeating the analysis using paralaxes from \textit{Hipparcos} \citep{Lee09} wields similar trends, but with larger errorbars.

A study presented in \citet{Huber17} shows that parallaxes derived from seismology are underestimated when compared to parallaxes from \textit{Gaia} Data Release 1 for subgiants. \citet{Huber17} used a larger sample of $\approx$ 1800 red giants to calculate the residual offset between radii from \textit{Gaia} DR1 with asteroseismic radii. However, \citet{Huber17} found no significant offset for more evolved stars when using a large sub-sample of RGB and red clump giants. Additionally, \citet{Zinn19} have found a zero-point offset in \textit{Gaia} Data Release 2 parallaxes using asteroseismic data of red giants in the \textit{Kepler} field. They used red giant branch stars from the APOKASC-2 catalogue to identify a color (and magnitude) dependent zero-point offset of $\sim 50$ $\mu\rm{as}$. Also, \citet{Khan19} found a similar zero-point offset in \textit{Gaia} DR2 parallaxes when using both \textit{Kepler} and K2 data combined with a grid-based method. More recently, \citet{Stassun21} used a sample of 158 eclipsing binaries with precise estimations of radii and effective temperature, which in turn allows precise values for bolometric luminosities via the Stefan-Boltzmann relation. They used their independently inferred parallaxes to compare with the parallaxes reported in \textit{Gaia} DR3, and have found a mean offset of $-37 \pm 20 \mu\rm{as}$, decreasing to $-15 \pm 18 \mu\rm{as}$ (\textit{Gaia}-EB) after applying corrections recommended by the \textit{Gaia} team. In this work, we only added offsets obtained following the work reported in \citet{Lindegren20}, where they presented an offset that is a function of magnitude G, ecliptic latitude and the effective wavenumber used in the astrometric solution. For our 9 stars, this offset in DR3 parallaxes is on average $-20$ $\mu\rm{as}$.

Moreover, one of the stars in our sample (KIC 10323222) has a measurement of its limb-darkened interferometric angular diameter \citep{Karo22}. A direct interferometric measure of stellar radii is of particular importance for establishing the accuracy of the inferred stellar parameters. For KIC 10323222, the stellar radius obtained by using the measured angular diameter is $9.796 \pm 0.137$ $\rm R_{\odot}$, and the stellar mass and surface temperature are $1.23 \pm 0.12$ $\rm M_{\odot}$ and $4568 \pm 42$ K, respectively \citep[see][]{Karo22}. The corresponding stellar radius obtained in this work is $\sim 5\%$ higher on average ($10.34 \pm 0.27$ $\rm R_{\odot}$), and was calculated using the scaling relations, where global seismic parameters are combined with spectroscopic estimations of temperatures (see equation \ref{seismicradii}). In another studies, \citet{Gaulme16} and \citet{Benba21} also used independent methods to obtain stellar radii in a group of red giants in eclipsing binaries. Their work indicates that seismic radii are $\sim 5\%$ too high for red giants. Additionally, \citet{Brogaard18} have partially revisited the sample of eclipsing binaries from \citet{Gaulme16} while considering corrections to $\Delta\nu$ scaling. They found a complex situation where some stars show agreement between the dynamical and corrected asteroseismic measures while others suggest significant overestimates of the asteroseismic measures. They could not pinpoint a simple explanation for this, but found indications of several potential problems, some theoretical, others observational \citep[see][for more information]{Brogaard18}. \citet{Kalli18} has also conducted an analysis of the six seismic binaries from \citet{Gaulme16} and applied probabilistic methods to re-examine the global oscillation parameters of the giants in the binary systems. They have found a good agreement between seismic and dynamic parameters when using a new nonlinear scaling relations for $\Delta\nu$ and $\nu_{\rm{max}}$.

\subsection{Results from grid-based pipelines}
\label{grid_results}

\begin{figure*}

 \includegraphics[width=\columnwidth]{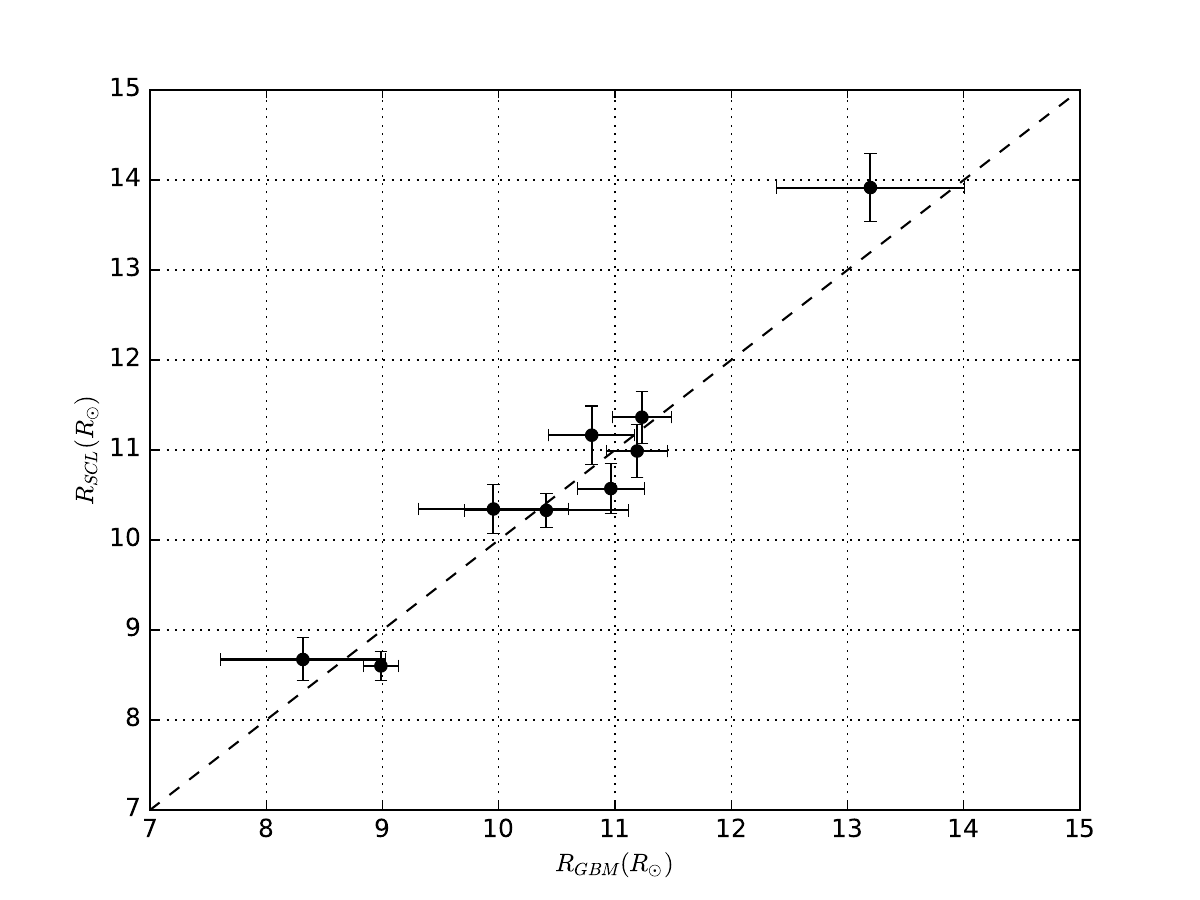}
 \includegraphics[width=\columnwidth]{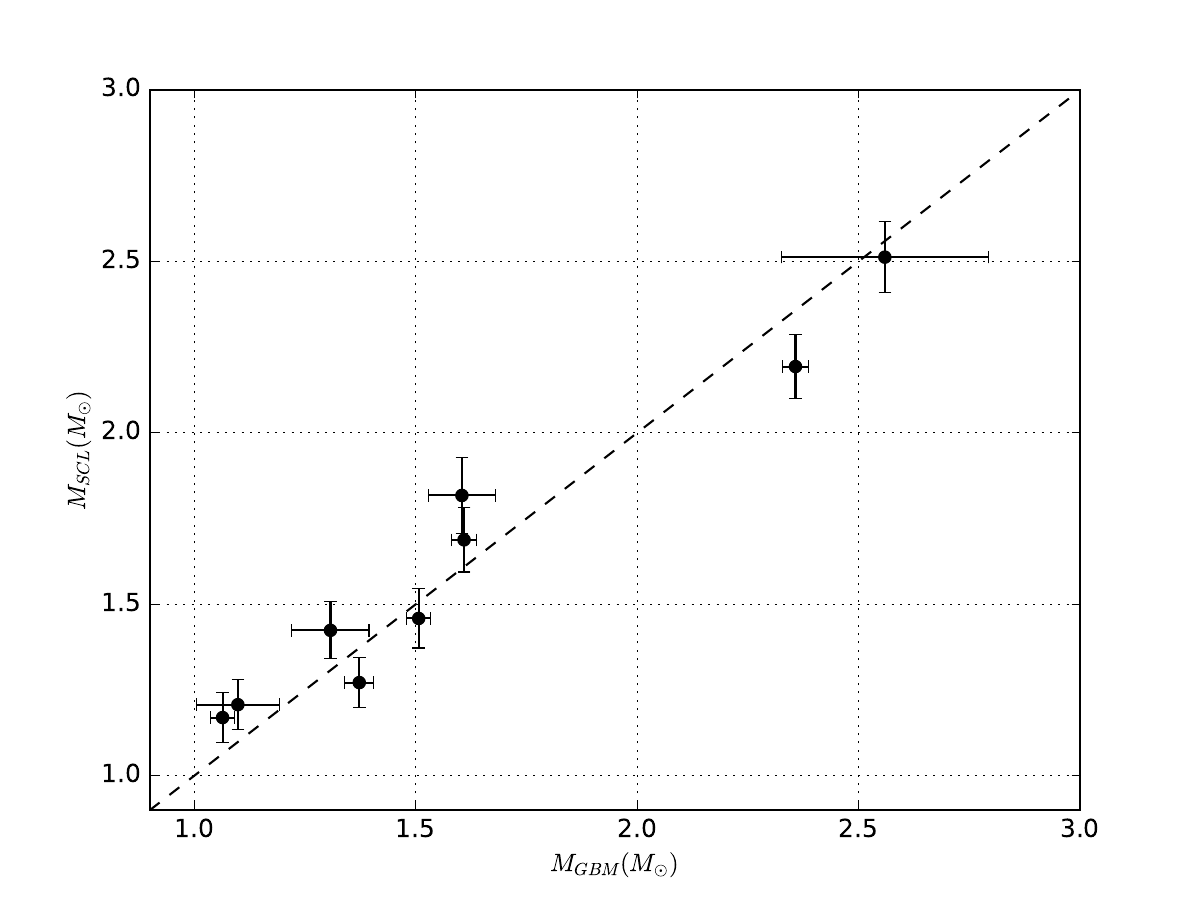}

 \caption{Left panel: Comparison between the radii obtained by grid based modelling (\textit{GBM}) pipeline and the radii obtained by using seismic data and the scaling relations (\textit{SCL}). Black dashed line represents the identity line. Right panel: same as the left panel, now using stellar masses instead.}

 \label{global_param}
\end{figure*}

\begin{figure}

 \includegraphics[width=\columnwidth]{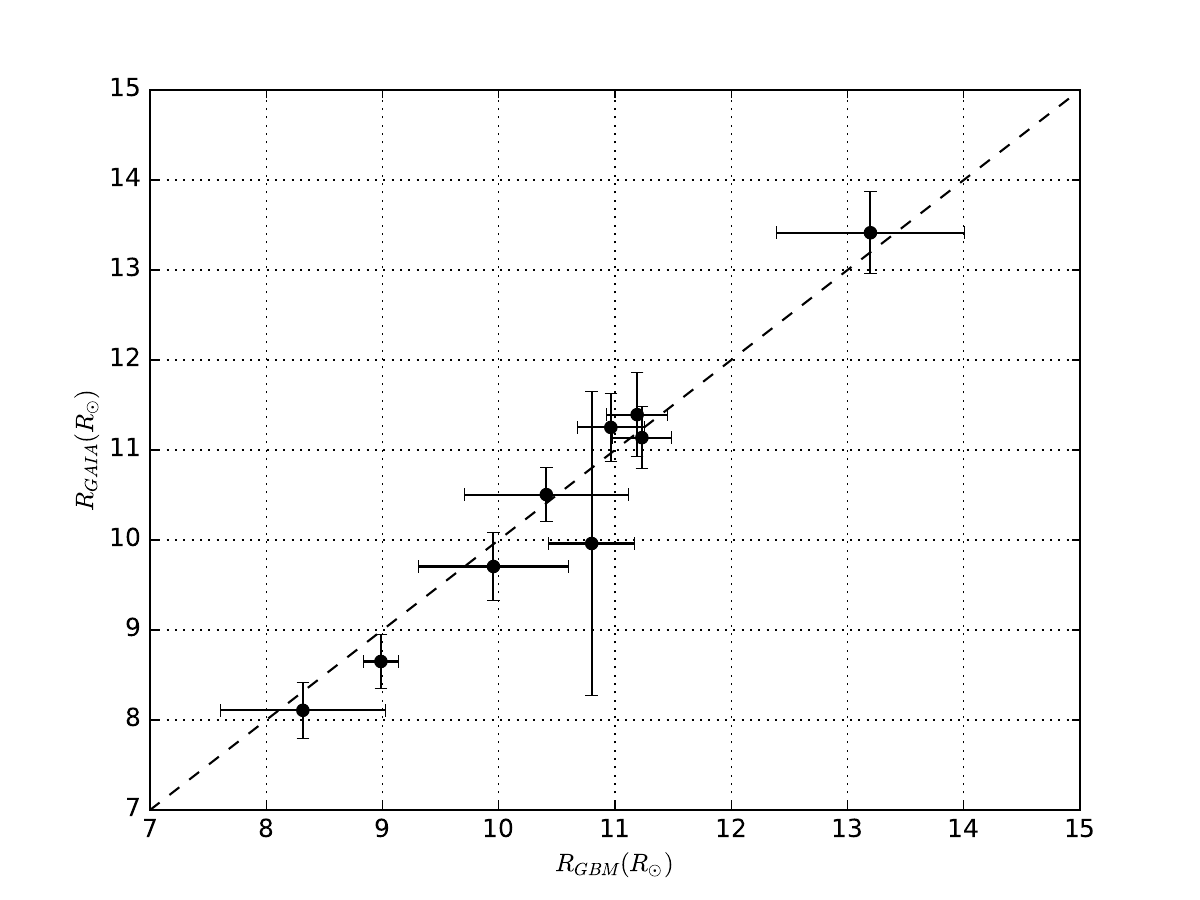}

 \caption{Comparison between stellar radii obtained by grid based modelling (\textit{GBM}) pipeline and the radii obtained by using \textit{Gaia} DR3 parallaxes. Black dashed line represents the identity line.}

 \label{gridR_vs_plx}
\end{figure}

\begin{figure}
 
 \includegraphics[width=\columnwidth]{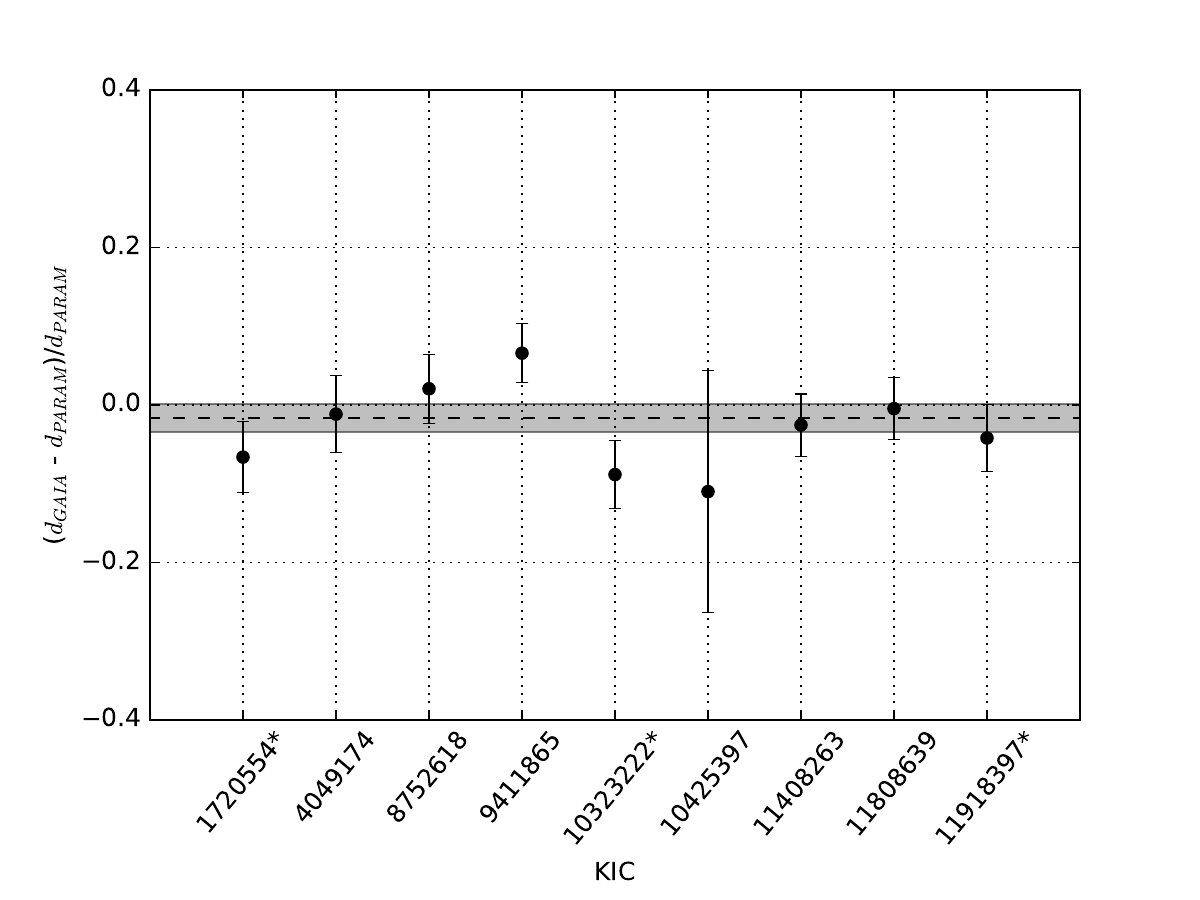}

 \caption{Residual difference between distances inferred from PARAM and distances obtained from \textit{Gaia} DR3 parallaxes. Distances from PARAM are estimated using radii obtained by grid-modelling (see Sect. \ref{grid} for more details), spectroscopic $T_{\rm eff}$ and Ks magnitudes. The weighted average of the residuals is -0.016 (black dashed line) with a statistical error of 0.017 (shaded area).}

 \label{param_dist}
\end{figure}


In this Section we infer stellar properties by making use of the following grid-based pipelines:
\begin{itemize}

\item The Pisa group adopted the SCEPtER pipeline \citep{Valle14, Valle15} coupled with a grid of stellar models computed as described in \citet{Dell12}. The grid contains models computed without core overshooting and does not account for mass loss on the RGB. Scaling relations were used to obtain the average asteroseismic parameters.

\item Non-canonical BaSTI \citep{Pietrinferni2004} isochrones with no mass loss and including effects of overshooting \citep[described in][]{SilvaAguirre13}. The pipeline applied is the BAyesian STellar Algorithm (BASTA), described in \citet{SilvaAguirre15, SilvaAguirre18}. Results have been computed using the solar values from OCT \citep{Hekker10} in the scaling relations. Corrections to the theoretical values of $\Delta\nu$ were also included.

\item PARAM pipeline (\citealt{daSilva06}, \citealt{Miglio13a}) using Bayesian statistics to obtain global stellar properties from a grid of models. Evolutionary tracks were computed with MESA \citep{Pax11}, and each of them has a precomputed set of frequencies that PARAM uses to calculate the large separation directly \citep{Rodrigues17}, avoiding the use of the $\Delta \nu$ scaling relation and the systematic effects associated with it \citep[see, e.g.,][]{White13}.

\item {BeSPP is a Bayesian tool to infer stellar properties from spectrophotometric and asteroseismic data. The grid of stellar models has been constructed using GARSTEC \citep{Weiss08}. The large frequency separation has been determined for each stellar model from the  slope of the radial $\ell=0$ orders around $\nu_{\rm max}$, as in \citet{White11}, i.e. no $\Delta\nu$ scaling relation is used.}

\item The AMS code makes use of two grids of stellar models. The first one is the Canonical BaSTI grid \citep{Pietrinferni2004}, with a range of masses going from 0.5 $M_{\odot}$ to 3.5 $M_{\odot}$ in steps of 0.05 $M_{\odot}$. The second grid was constructed using MESA and covered the same selection of initial metallicities and helium abundances as the BaSTI grid but with a different mass range, from 0.8 $M_{\odot}$ to 1.8 $M_{\odot}$ in steps of 0.01 $M_{\odot}$. AMS uses two methods to fit stellar models to the observed parameters. They are independent implementations of the likelihood distribution method described by \citet{Basu10} and the SEEK method described in \citet{Quirion10}. More details about AMS can be found in \citet{Hekker14}.

\end{itemize}

The main procedure is to run the pipelines using the following parameters as input constraints: $\nu_{\rm{max}}$, $\Delta\nu$, $T_{\rm{eff}}$ and [Fe/H]. In return, the codes obtain a comprehensive list of stellar attributes, including but not limited to: masses, radii, surface gravities, ages, and densities (and their associated uncertainties). In some occasions, the asymptotic period spacing, when it could be measured (see Table \ref{seismic_table}), was used to give some insights on the evolutionary status. The pipelines that calculate $\Delta\nu$ directly from model frequencies are calibrated to mitigate the impact caused by surface effects.
 
We compared stellar properties obtained by the models with their respective values obtained using scaling relations. We decided to select one pipeline to ensure homogeneous input physics. In order to do so, we took results from all pipelines and computed the median value for each global stellar properties. Our results show that the AMS pipeline has results closest to the median, so we choose to use AMS values as the reference global stellar properties from grid-based modelling (GBM). Errors were calculated from the quadratic sum of uncertainties of different results.

The left panel of Figure \ref{global_param} shows a comparison between the stellar radii from grid based modeling and radii estimated using scaling relations (SCL). The weighted average of the residual difference is 0.009 with a statistical error of 0.013. In Figure \ref{gridR_vs_plx}, we show the a similar comparison between radii obtained from grid based modelling and radii obtained from \textit{Gaia} DR3 parallaxes. The weighted average of the residuals is 0.004 with a statistical error of 0.009. As for the stellar masses, the right panel of Figure \ref{global_param}  show the comparison between stellar masses from grid-based modelling and masses derived from scaling relations. The weighted average of the residuals is -0.012 with a statistical error of 0.026. It is also important to note that the grid results for the three ascending red giant stars have systematically lower values of mass when compared with the SCL values. This can be partially explained by the biases in the $\Delta\nu$ scaling (see Figure \ref{Dnu_scaling}).

The PARAM pipeline can estimate distances and extinctions by combining photometric data in several observed bands. However, most stars in our sample have either large uncertainties in observed magnitudes or contain flags of poor photometric quality, mostly due to saturation. Therefore, magnitudes are, in most cases, not accurate enough to be able to estimate extinctions using PARAM. To conduct a better comparison in terms of distances, we combined radii obtained from PARAM with spectroscopic temperatures and magnitudes (with extinctions from dustmaps) in Ks band to estimate a set of distances reffered to as $d_{\rm PARAM}$. Figure \ref{param_dist} shows a comparison between $d_{\rm PARAM}$ and distances derived from \textit{Gaia} DR3 parallaxes. Residual differences show a good agreement, with a weighted average of -0.016 and a statistical uncertainty of 0.017. To check the effects of using a different photometric band, we switched Ks band for V band in the calculations of $d_{\rm PARAM}$, and compared it to \textit{Gaia} DR3 parallaxes once again. By doing this, the residual differences increase slightly to $0.025 \pm 0.020$.

\section{Discussion of key chemical elements}
\label{discu}

\begin{figure*}
 
 \includegraphics[width=\columnwidth]{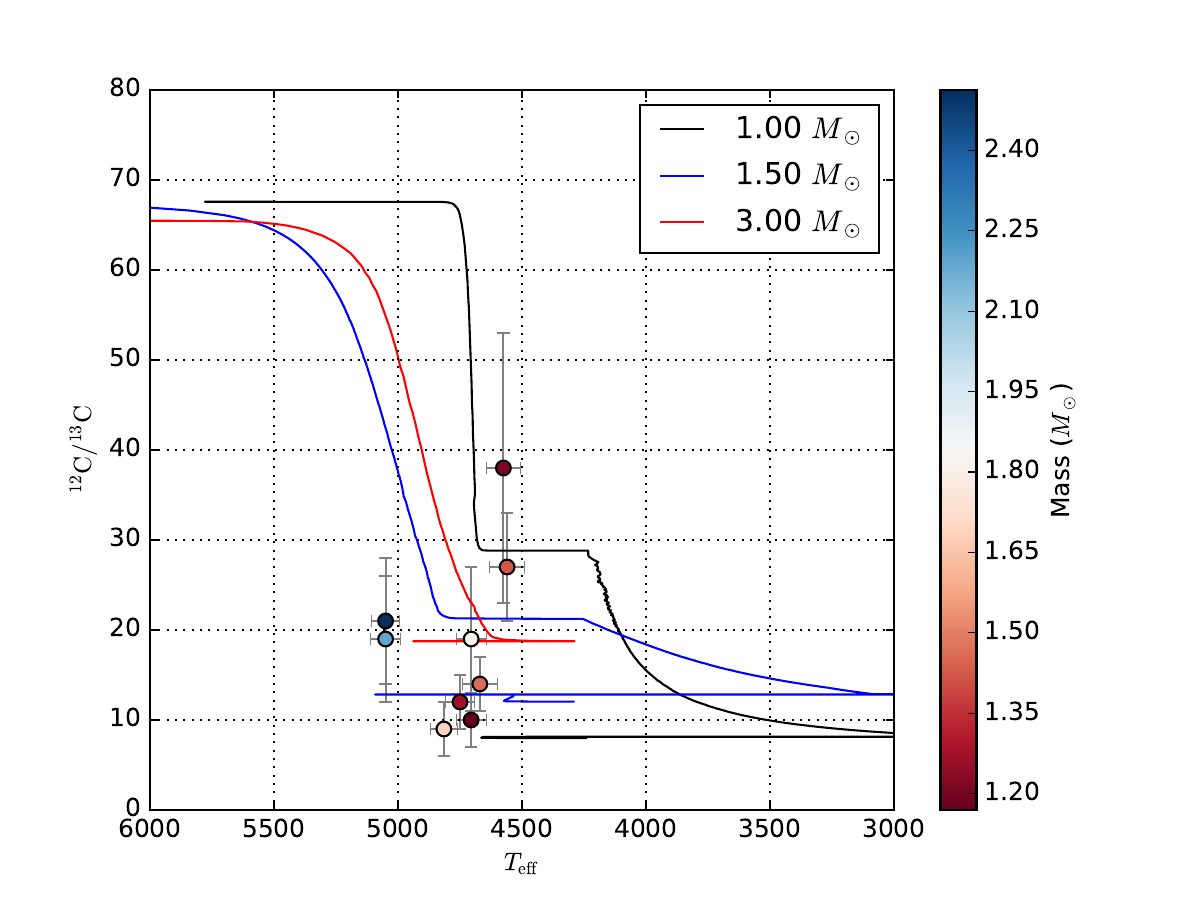}
 \includegraphics[width=\columnwidth]{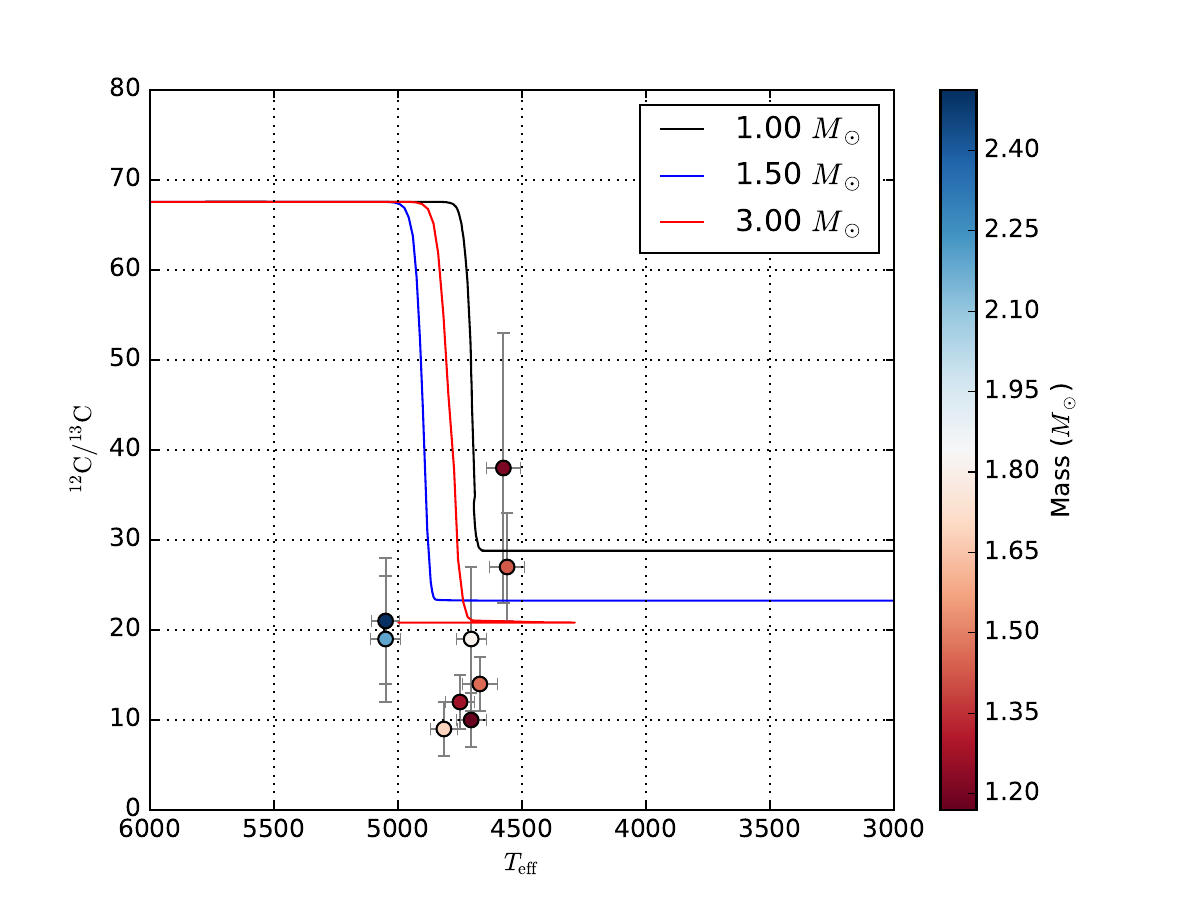} 

 \caption{The theoretical evolution of carbon isotopic ratio $^{12}$C/$^{13}$C at the stellar surface for models with M=1.0, 1.5 and 3.0 $\rm M_{\odot}$, including the effects of thermohaline instability and rotation-induced mixing (left panel) and following the standard prescriptions (right panel). All stellar tracks have solar metallicity. Circles represent our \textit{Kepler} red giant stars.}

 \label{C_ratio}
\end{figure*}

\begin{figure*}

 \includegraphics[width=\columnwidth]{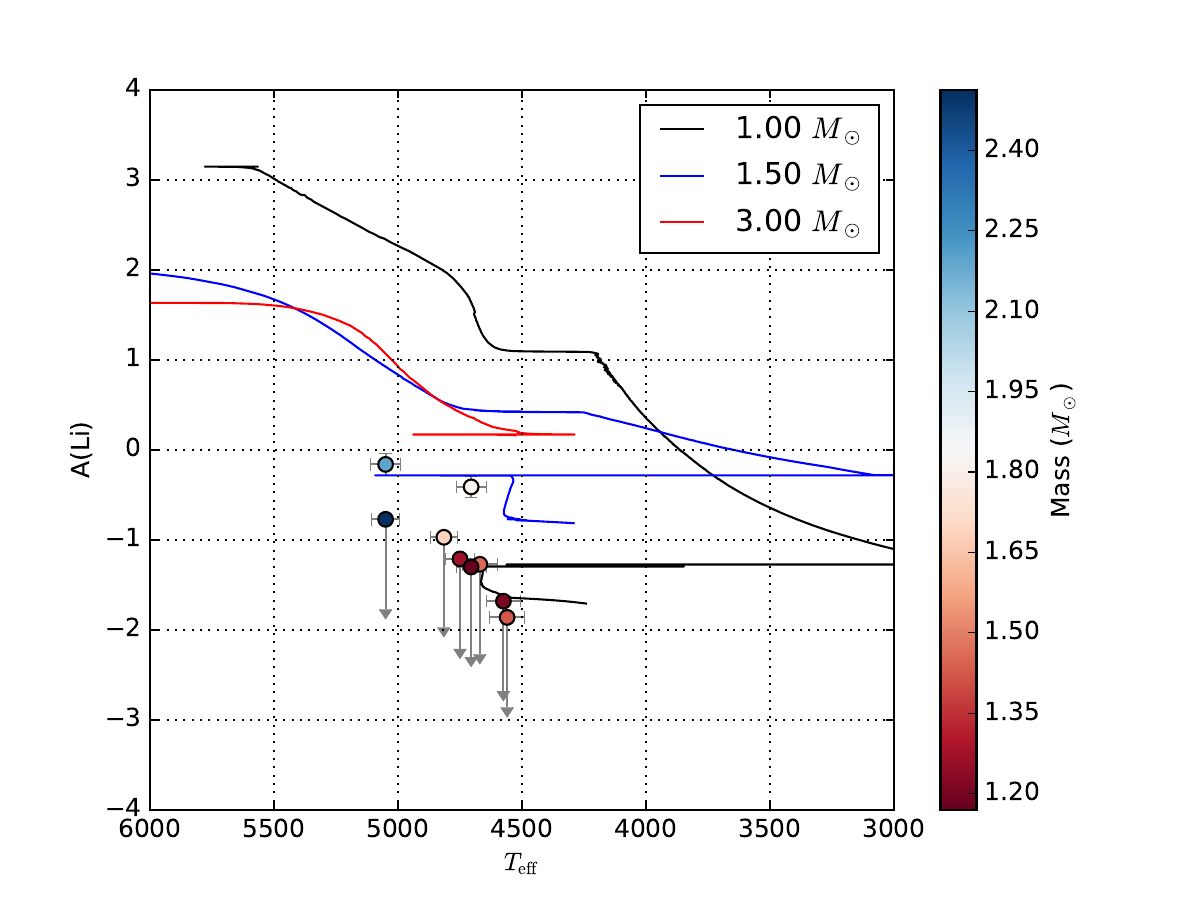}
 \includegraphics[width=\columnwidth]{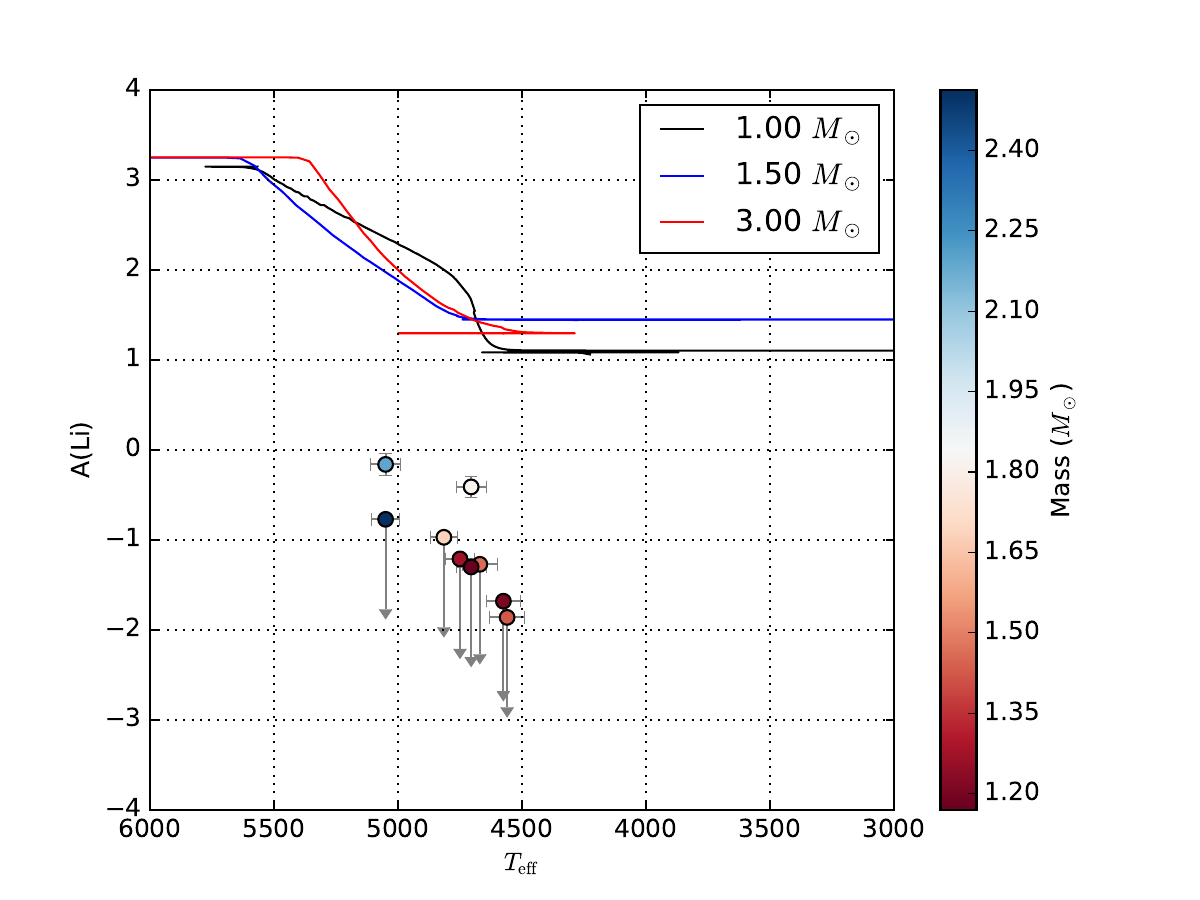} 

 \caption{Synthetic evolution of Lithium abundances A(Li) at the stellar surface as a function of $T_{\rm eff}$. We show models with M=1.0, 1.5 and 3.0 $\rm M_{\odot}$, including the effects of thermohaline instability and rotation-induced mixing (left panel) and following the standard prescriptions (right panel). All stellar tracks have solar metallicity. Circles represent our \textit{Kepler} red giant stars, while downward arrows represent stars that only have an upper limit in A(Li).}

 \label{A_li}
\end{figure*}

\subsection{Surface Lithium and Carbon isotopic ratio}

In order to compare the observed abundances with our theoretical predictions, we obtained stellar evolution models computed with the STAREVOL code \citep{Lagarde12, Amard19}, including the effects of rotation-induced mixing and thermohaline instability all along the stellar evolution \citep{Lagarde12}. These models reproduced spectroscopic data in open clusters and field stars \citep[e.g.][]{Charb10, Charb20, Magrini21}. Although its efficiency is still discussed in the literature, thermohaline mixing seems to be the main physical process governing the surface abundances of C, N, and Li for stars more evolved than the bump luminosity on the red giant branch. \cite{Charb10} showed that its efficiency is decreasing with increasing initial stellar mass. On the other hand, rotation-induced mixing modifies the internal chemical profiles during the main sequence \citep[e.g.][]{Pala03, Pala06}. Its effects on the surface abundances are revealed only after the first dredge-up at the base of the RGB. Including the effects of these extra-mixing events in theoretical models is a crucial step to provide a more comprehensive understanding of the surface chemical patterns observed in low-mass red-giant stars, as each mechanism alone cannot fully explain the observed abundance variations.

According to previous studies, we distinguish three regimes related to initial stellar mass: low-mass stars (M $\leq$ 1.8 M$_{\odot}$), where thermohaline mixing is the main process that changes the surface abundances compared to rotation-induced mixing, since the latter is weaker in lower-mass stars due to their slower rotation rates. For intermediate-mass stars (1.8 $\leq$ M $\leq$ 2.2 M$_{\odot}$) both mechanisms play an equivalent role to change surface abundances. And finally, for more massive stars (M $>$ 2.2 M$_{\odot}$) thermohaline mixing plays no role because these stars ignite central Helium-burning before reaching the RGB bump. As a result, thermohaline mixing does not occur in these stars, only rotation is expected to have an impact on surface abundances. 

In addition to carbon isotopic ratio, we also obtained estimations for the abundance of surface lithium, where we addopted non-LTE values for the latter.\footnote{Lithium abundance is defined here as $A(Li) = \log \left(\frac{X(Li)}{X(H)} \frac{A_{\rm{H}}}{A_{\rm{Li}}}\right)+12$, where $X(\rm{Li})$ is the lithium mass fraction.} Figure \ref{C_ratio} shows the theoretical evolution of carbon isotopic ratio for three models at 1.0 $\rm M_{\odot}$, 1.5 $\rm M_{\odot}$ and 3.0 $\rm M_{\odot}$, all at solar metallicity, while Figure \ref{A_li} shows the theoretical evolution of Lithium for the same stellar evolution models. They show the evolution from the zero-age main sequence to the early AGB phase. Non-standard models have an initial rotation velocity ($\rm{V_{ZAMS}}$) of 110 km/s. We make use of these figures to compare theoretical predictions with our observations.

\subsection{Low-mass stars ($M < 1.8$ $ M_{\odot}$)}
\begin{itemize}

\item KIC1720554 ($M = 1.21 \pm 0.07$ M$_{\odot}$) is a low-mass RGB star according to the period spacing ($\Delta\Pi$ = 66 s). This is corroborated by the measured value of $^{12}$C/$^{13}$C (38$\pm$ 15) and the position of this star in the HR diagram. Those properties are a strong evidence that this is a RGB star, not evolved enough to undergo thermohaline mixing. The surface lithium abundance in KIC 1720554 is estimated to be lower than -1.68, and such low values of A(Li) suggests the occurence of extra mixing events. It is known that initial rotation rates can have a strong inpact on the depletion of lithium during the main sequence. Therefore,  rotation-induced mixing could be responsible for the current low amount of lithium. Investigating the specific mixing mechanisms and their impact on the evolution of this star would require further study.

\item KIC10323222 ($M = 1.42 \pm 0.08$ M$_{\odot}$) does not have a measured value for the period spacing. However, results from seismology show that this is a low-mass star, while spectroscopy wields a high $^{12}$C/$^{13}$C (27$\pm$ 6) at the surface. Combining this information with its position in the HR diagram, there is a strong argument that this is a star on the RGB, which supports our early assumption (see Sect. \ref{extract}). The lithium abundance in KIC10323222 is estimated to be lower than -1.86. This is similar to the situation of KIC1720554, where such low amount of surface lithium is evidence of extra mixing, with rotation based mixing being a strong cadidate. A more detailed analysis would be required to better understand the mechanisms involved.

\item KIC4049174, KIC8752618, KIC10425397 and KIC11408263 are a group of low-mass stars (between $\sim$ 1.17 and 1.69 M$_{\odot}$) that share similar characteristics in regards to chemical profile. The period spacing of mixed modes ($\Delta\Pi$ $\sim$ 300s) indicates status of clump stars while the low surface carbon isotopic ratio (lower than 14) and lithium abundances (upper limit of -1.13) confirms this evolutionary state by comparison with stellar evolution models including thermohaline mixing (see Figures \ref{C_ratio} and \ref{A_li}). The right panel of Fig \ref{C_ratio} shows that standard models are not able to reproduce such low values of $^{12}$C/$^{13}$C and lithium. 

\end{itemize}

\subsection{Intermediate-mass stars ($1.8 < M (M_\odot) < 2.2$)}

\begin{itemize}

\item KIC 11918397 ($M = 1.82 \pm 0.11 M_{\odot}$) is an intermediate-mass star with low-metallicity ([Fe/H]=-0.21$\pm$ 0.10). Although the period spacing could not be measured for this star, its chemical properties ($^{12}$C/$^{13}$C=19$\pm$ 8 and A(Li) = $-0.41 \pm 0.12$) seem to corroborate the assumption that this star is still in the RGB. 

\item According to the asymptotic period spacing ($\Delta\Pi$ = 226.9s), KIC11808639 ($M=2.19 \pm$ 0.09 $M_{\odot}$) is a red giant in the secondary clump. The measured surface chemical properties (A(Li) = $-0.16\pm 12$ and $^{12}$C/$^{13}$C=19$\pm$ 7) are consistent with this evolutionary state, and are explained with stellar models including rotation and thermohaline mixing (Figures \ref{C_ratio} and \ref{A_li}). 

\end{itemize}

\subsection{More massive stars ($M > 2.2$ $M_{\odot}$)}

In more massive stars, theoretical models do not predict changes in the surface chemical properties during the red giant branch. According to stellar evolution models, RGB and clump stars show the same surface $^{12}$C/$^{13}$C and lithium abundances. For this mass range, rotation-induced mixing plays a dominant role and changes the surface properties of early-RGB stars. 

\begin{itemize}

\item KIC9411865 ($M=2.51 \pm $0.10 M$_{\odot}$) is classified as being in the secondary clump, according to the period spacing. Although the upper limit of lithium (A(Li) < -0.77) abundance does not give additional constraints, this evolutionary state is consistent with its measured carbon isotopic rate $^{12}$C/$^{13}$C $= 21 \pm 7$.
 
\end{itemize}

\section{Conclusions} 

We report results for various seismic tests and analysis using a group of nine nearby red giants observed by \textit{Kepler}. The stars in our sample have accurate values of parallaxes obtained by \textit{Gaia} DR3. We combined seismic data and detailed spectroscopic parameters to infer global stellar properties. Results from both direct aplication of seismic scaling relations and predictions from grid-based models indicate that our sample is in the mass range where two non-canonical mixing processes have been suggested to play a significant part in explaining observed values of surface chemical abundances. The results we obtained in this work are summarised as follows:

\begin{itemize}

\item In the context of constrained spectroscopic analysis using seismic $\log g$, we have ascertained that using excitation balance yields slightly larger values of effective temperature for hotter ($T_{\rm eff} \geq 5000$ K) stars. Our results using iron ionization balance and seismic $\log g$ yield values of $T_{\rm eff}$ which are in good agreement with estimations of photometric $T_{\rm eff}$. 

\item We compared differences between distances derived from \textit{Gaia} DR3 parallaxes with distances inferred by using seismic parameters and scaling relations at face value, obtaining a weighted average of the residual differences equal to $-0.0134 \pm 0.0150$. Although this suggests a small difference between \textit{Gaia} and seismic distances, we emphasize that the size of uncertainties are such that this result is statistically consistent with zero. We also applied a simple correction to the values of $\Delta\nu$ of 3 RGB stars based on a grid of models (Figure \ref{Dnu_scaling}) in order to test how it would affect the comparison with seismic distances. The effects of this correction in $\Delta\nu$ is an increase in the previous offset in distances by $\sim 2\%$. Finally, it is important to stress that reddening is a fundamental source of uncertainty, since it has a significant effect on the calculations of seismic distances. Adopting 2MASS Ks magnitudes appears to significantly mitigate the effects of reddening (see appendix \ref{appendix_dust}). It is also important to note that derivations of seismic distances (equation \ref{seismodist}) are affected by the chosen scale of effective temperature, turning $T_{\rm eff}$ into a source of systematic error. 

\item We used radii inferred by grid-models from PARAM, along with spectroscopic $T_{\rm eff}$ and magnitudes, to compute a set of distances to be compared with distances derived from \textit{Gaia} DR3 parallaxes. Results show an excellent agreement, however the small amount of stars in our sample and the size of uncertainties are such that a larger number of stars would be necessary to produce a more stringent result.

\item We calculated residual differences in stellar radii obtained by combining parallaxes from \textit{Gaia} DR3 with radii inferred directly from scaling relations. The weighted average indicates that radii derived from \textit{Gaia} are lower than the asteroseismic radii by $\sim 1\%$. We note again that, given the size of uncertainties, those results are statistically consistent with zero. \citet{Huber17} used a larger sample of $\approx$ 1800 red giants to calculate the residual offset between radii from \textit{Gaia} Data Release 1 with asteroseismic radii, and have found no significant offset for low luminosity red giants. When grid-based modelling is applied to our stars, there is a small offset between the radii derived from measured parallaxes and the grid-based radii. Radii derived from grid-based modelling are $\sim 0.4\%$ higher than radii derived from \textit{Gaia} DR3 parallaxes. It is important to note that \citet{Gaulme16} and \citet{Benba21} used a sample of red giant stars in eclipsing binaries, taking advantage of independent methods to obtain masses and radii. They concluded that seismic radii based on seismic relations at face value are $\sim 5\%$ too high for red giants. Additionally, \citet{Brogaard18} added corrections to $\Delta\nu$ scaling while studying the sample of \citet{Gaulme16}. They found that some stars show agreement between the dynamical and corrected asteroseismic measures, while others suggest significant overestimates of the asteroseismic measures. They could not find a simple explanations for such complex situation, but found indications of several potential problems \citep[see][for more information]{Brogaard18}. Finally, \citet{Zinn19} and \citet{Khan19} used asteroseismic data of red giants in the Kepler field to find a zero-point offset in \textit{Gaia} DR2 parallaxes of a few tens of $\mu\rm{as}$. If a similar offset is applied to the stars of our sample, the effects in the residual differences of distances would be of the order of 0.5$\%$ or less.

\item For stellar masses, we compared values obtained using different combinations of seismic and non-seismic constraints. We obtained a typical mass uncertainty of $\backsim$ 8$\%$ when combining $\nu_{\rm{max}}, \Delta\nu$ and spectroscopic $T_{\rm{eff}}$, $\backsim$ 23$\%$ when using $\Delta\nu$ and the radius obtained from \textit{Gaia} DR3 parallaxes and, finally, $\backsim$ 14$\%$ when using the combination of $\nu_{\rm{max}}$, spectroscopic $T_{\rm{eff}}$ and radii derived from \textit{Gaia} DR3 parallaxes. These results suggest that the size of the uncertainties in stellar masses obtained with the combination of radii derived from \textit{Gaia} DR3 parallaxes with at least one global seismic parameter (either $\Delta\nu$ or $\nu_{\rm{max}} + T_{\rm{eff}}$, see Equations \ref{mass1} and \ref{mass2} ), will be similar to the size of uncertainties in stellar masses derived by combining $\nu_{\rm{max}}, \Delta\nu$ and spectroscopic $T_{\rm{eff}}$ (Equation \ref{seismicmass}).

\item Our precise estimations of stellar masses (with uncertainties smaller than $10\%$) with a solid proxy for evolutionary stage in the form of $\Delta\Pi$ (distinguishing RGB from core He-burning) make it clear that the stars in our sample fall in the mass range where it is claimed that thermohaline instability coupled with the effects of rotation is the most efficient transport processes for chemical elements. Therefore, we compared the observed surface abundances of lithium and carbon with predictions made by models with and without extra-mixing processes. Models that do not include the effects of extra-mixing processes fail to explain the observed surface abundances of lithium and carbon isotopic ratio. Our results reinforce the idea that the effects of both rotation-induced mixing and thermohaline instability explain the spectroscopic observations of the aforementioned elements in low- and intermediate-mass red giants stars. It is important to point out, however, the rather loose constraints that the abundances of these key elements impose on models that account for rotational mixing. The initial rotation velocity in particular, despite its importance for the models, is an arbitrary parameter that can be adjusted, to some extent, in order to obtain a better agreement with the observed data \citep[for a detailed discussion see][Figure 6]{Lagarde15}. 

\end{itemize}

We expect to better quantify the trends presented here by using individual mode frequency modelling combined with \textit{Gaia} DR3 parallaxes. We also seek to increase the number of stars in the sample in order to conduct more precise scaling relation tests. 

\section{DATA AVAILABILITY AND ONLINE MATERIAL}
All the data underlying this article is fully available under reasonable request to the corresponding author. \footnote{hugo.pesquisa@gmail.com}

\subsection*{ACKNOWLEDGMENTS}

H.R.C. acknowledge the Brazilian research funding agency CNPq for financial support. This work has made use of data from the European Space Agency (ESA) mission {\it Gaia} (\url{http://www.cosmos.esa.int/gaia}), processed by the {\it Gaia} Data Processing and Analysis Consortium (DPAC, \url{http://www.cosmos.esa.int/web/gaia/dpac/consortium}). Funding for the DPAC has been provided by national institutions, in particular the institutions participating in the {\it Gaia} Multilateral Agreement. T.M. acknowledges financial support from Belspo for contract PRODEX "PLATO mission development". N. L. acknowledges financial support from "Programme National de Physique Stellaire" (PNPS) and from the "Programme National Cosmology et Galaxies (PNCG)" of CNRS/INSU, France. R.A.G. acknowledges the support from the PLATO Centre National d'\'Etudes Spatiales grant. Finally, the authors thank Remi Cabanac, former Director of the Pic du Midi Observatory, for the quick allocation of the observing time.

\clearpage
\newpage

\begin{sidewaystable}
\begin{tiny}
\begin{center}
\caption{Chemical abundances for several elements. For the lithium abundances, values denote only upper limits for all but two stars. The NLTE values for Li are to be preferred over LTE. The oxygen abundance is based on [O I] 6300 \AA.}
\begin{tabular}{c c c c c c c c c c c c c c c c c c c}
\hline
KIC & $^{12}$C/$^{13}$C & A(Li)(LTE) & A(Li)(NLTE) & $\sigma$ A(Li) & [C/Fe] & [N/Fe] & [O/Fe] & [Na/Fe] & [Ni/Fe] & [Mg/Fe] & [Al/Fe] & [Si/Fe] & [Ca/Fe] & [Sc/Fe]  & [Ti/Fe] & [Cr/Fe] & [Co/Fe] & [Ba/Fe]\\
\hline
1720554 & $38\pm 15$ & $<-1.89$ & $<-1.68$ & $...$ & $ 0.04\pm 0.09$ & $0.17\pm 0.13$ & $ 0.08\pm 0.10$ & $0.17\pm 0.08$ & $0.04\pm 0.08$ & $ 0.22\pm 0.08$ & $0.32\pm 0.13$ & $0.20\pm 0.08$ & $ 0.13\pm 0.06$ & $ 0.09\pm 0.08$ & $ 0.11\pm 0.11$ & $0.00\pm 0.06$ & $ 0.26\pm 0.08$ & $ 0.00\pm 0.08$ \\
4049174 & $14\pm  3$ & $<-1.49$ & $<-1.27$ & $...$ & $-0.12\pm 0.09$ & $0.40\pm 0.13$ & $-0.03\pm 0.10$ & $0.22\pm 0.06$ & $0.04\pm 0.09$ & $ 0.07\pm 0.08$ & $0.14\pm 0.08$ & $0.19\pm 0.06$ & $-0.02\pm 0.08$ & $ 0.03\pm 0.08$ & $-0.04\pm 0.11$ & $-0.05\pm 0.10$ & $ 0.21\pm 0.08$ & $ 0.00\pm 0.09$ \\
8752618 & $12\pm  3$ & $<-1.39$ & $<-1.21$ & $...$ & $-0.17\pm 0.09$ & $0.26\pm 0.13$ & $ -0.05\pm 0.10$ & $0.13\pm 0.06$ & $0.02\pm 0.07$ & $ 0.10\pm 0.08$ & $0.15\pm 0.08$ & $0.21\pm 0.10$ & $ 0.04\pm 0.06$ & $ -0.05\pm 0.08$ & $ -0.01\pm 0.09$ & $-0.04\pm 0.06$ & $ 0.15\pm 0.08$ & $ 0.08\pm 0.09$ \\
9411865 & $21\pm  7$ & $<-0.89$ & $<-0.77$ & $...$ & $-0.28\pm 0.09$ & $0.42\pm 0.13$ & $-0.19\pm 0.10$ & $0.17\pm 0.05$ & $-0.01\pm 0.05$ & $0.02\pm 0.08$ & $0.04\pm 0.05$ & $0.14\pm 0.08$ & $ 0.06\pm 0.05$ & $-0.10\pm 0.08$ & $-0.07\pm 0.08$ & $-0.03\pm 0.07$ & $-0.03\pm 0.08$ & $ 0.31\pm 0.10$ \\
10323222 & $27\pm  6$ & $<-2.09$ & $<-1.86$ & $...$ & $-0.02\pm 0.09$ & $0.38\pm 0.13$ & $ 0.02\pm 0.10$ & $0.24\pm 0.06$ & $0.05\pm 0.09$ & $ 0.07\pm 0.08$ & $0.21\pm 0.10$ & $0.19\pm 0.06$ & $ 0.05\pm 0.08$ & $ 0.04\pm 0.08$ & $ 0.01\pm 0.11$ & $-0.06\pm 0.05$ & $ 0.29\pm 0.08$ & $-0.07\pm 0.09$ \\
10425397 & $10\pm  3$ & $<-1.49$ & $<-1.30$ & $...$ & $-0.08\pm 0.09$ & $0.28\pm 0.13$ & $ 0.04\pm 0.10$ & $0.08\pm 0.06$ & $0.02\pm 0.07$ & $ 0.11\pm 0.08$ & $0.15\pm 0.07$ & $0.18\pm 0.07$ & $ 0.02\pm 0.06$ & $ ---$ & $-0.06\pm 0.09$ & $-0.09\pm 0.05$ & $ 0.14\pm 0.08$ & $ 0.09\pm 0.08$ \\
11408263 & $ 9\pm  3$ & $<-1.13$ & $<-0.97$ & $...$ & $-0.18\pm 0.09$ & $0.33\pm 0.13$ & $ 0.09\pm 0.10$ & $0.08\pm 0.05$ & $0.01\pm 0.05$ & $ 0.11\pm 0.08$ & $0.14\pm 0.07$ & $0.17\pm 0.09$ & $ 0.04\pm 0.05$ & $ -0.01\pm 0.08$ & $-0.04\pm 0.09$ & $-0.07\pm 0.05$ & $ 0.06\pm 0.08$ & $ 0.07\pm 0.09$ \\
11808639 & $19\pm  7$ & $-0.27$ & $-0.16$ & $0.12$ & $-0.26\pm 0.09$ & $0.33\pm 0.13$ & $-0.11\pm 0.10$ & $0.15\pm 0.05$ & $-0.04\pm 0.05$ & $0.02\pm 0.08$ & $0.03\pm 0.06$ & $0.100\pm 0.08$ & $ 0.05\pm 0.05$ & $-0.12\pm 0.09$ & $-0.09\pm 0.08$ & $-0.04\pm 0.08$ & $-0.02\pm 0.08$ & $ 0.28\pm 0.10$ \\
11918397 & $19\pm  8$ & $-0.59$ & $-0.41$ & $0.12$ & $-0.20\pm 0.09$ & $0.25\pm 0.13$ & $ 0.06\pm 0.10$ & $0.09\pm 0.06$ & $-0.03\pm 0.07$ & $ 0.09\pm 0.08$ & $0.12\pm 0.07$ & $0.15\pm 0.08$ & $ 0.06\pm 0.06$ & $ -0.01\pm 0.08$ & $ 0.00\pm 0.09$ & $ 0.00\pm 0.08$ & $ 0.01\pm 0.08$ & $ 0.32\pm 0.08$ \\
\hline
\end{tabular}
\label{chemical1}
\end{center}
\end{tiny}
\end{sidewaystable}
 
\begin{sidewaystable}
\begin{center}
\caption{Stellar magnitudes and derived properties for the stars in our sample.}
\begin{tabular}{c c c c}
\hline
KIC & \textit{Kepler} magnitude & 2MASS Ks magnitude & d (DR3)\\
  &  mag & mag & pc \\
\hline
1720554  & 6.525 & $4.154 \pm 0.018$ & $105.19 \pm 0.74$\\
4049174  & 6.279 & $4.081 \pm 0.036$ & $144.84 \pm 1.41$\\
8752618  & 6.441 & $4.299 \pm 0.015$ & $159.80 \pm 2.21$\\
9411865  & 6.114 & $4.135 \pm 0.017$ & $143.96 \pm 1.43$\\
10323222  & 6.725 & $4.318 \pm 0.017$ & $135.63 \pm 1.22$\\
10425397  & 6.449 & $4.507 \pm 0.362$ & $154.28 \pm 1.52$\\
11408263  & 6.307 & $4.091 \pm 0.018$ & $144.95 \pm 1.92$\\
11808639  & 6.229 & $4.234 \pm 0.036$ & $123.33 \pm 1.02$\\
11918397  & 6.235 & $3.963 \pm 0.016$ & $161.88 \pm 1.78$\\  
\hline
\end{tabular}
\label{derived_table}
\end{center}
\end{sidewaystable}


\newpage
\appendix
\section{Effects of different dust maps and magnitudes}
\label{appendix_dust}

The dust map and photometric bandpass were chosen such as to mitigate the impact of reddening on the determination of seismic distances. We tested a combination of two dust maps with four bands of magnitudes and how they would impact the calculations of relative differences between seismic distances and distances derived from \textit{Gaia } DR3 parallaxes. To estimate the effects of reddening we used the 3D dust maps from \citet{Drimmel03}, as implemented in the mwdust package \citep{Bovy16}. We also used the updated version of the dust map described in \citet{Green18}. We retrieved apparent magnitudes in the V band and in three near infrared JHKs magnitudes from 2MASS, where the effects of reddening have less impact. 

The effects of using different dust maps are more clear when using visual band (Figure \ref{res_dist_Vmag}) to calculate seismic distances, with the weighted average of the residuals going from $0.0354 \pm 0.0202$ when using data from \citet{Green18} to $0.0828 \pm 0.0257$ when using data from \citet{Drimmel03}. 

Meanwhile, the effects of using the two dust maps are much less significant when using the magnitudes from the 2MASS JHKs bands. We have adopted magnitudes from the 2MASS Ks band due to small values of uncertainties in apparent magnitudes when compared with J and H bands. We also decided to use reddening effects obtained by the dust maps presented in \citet{Green18}. 

\begin{figure*}
 
 \includegraphics[width=\columnwidth]{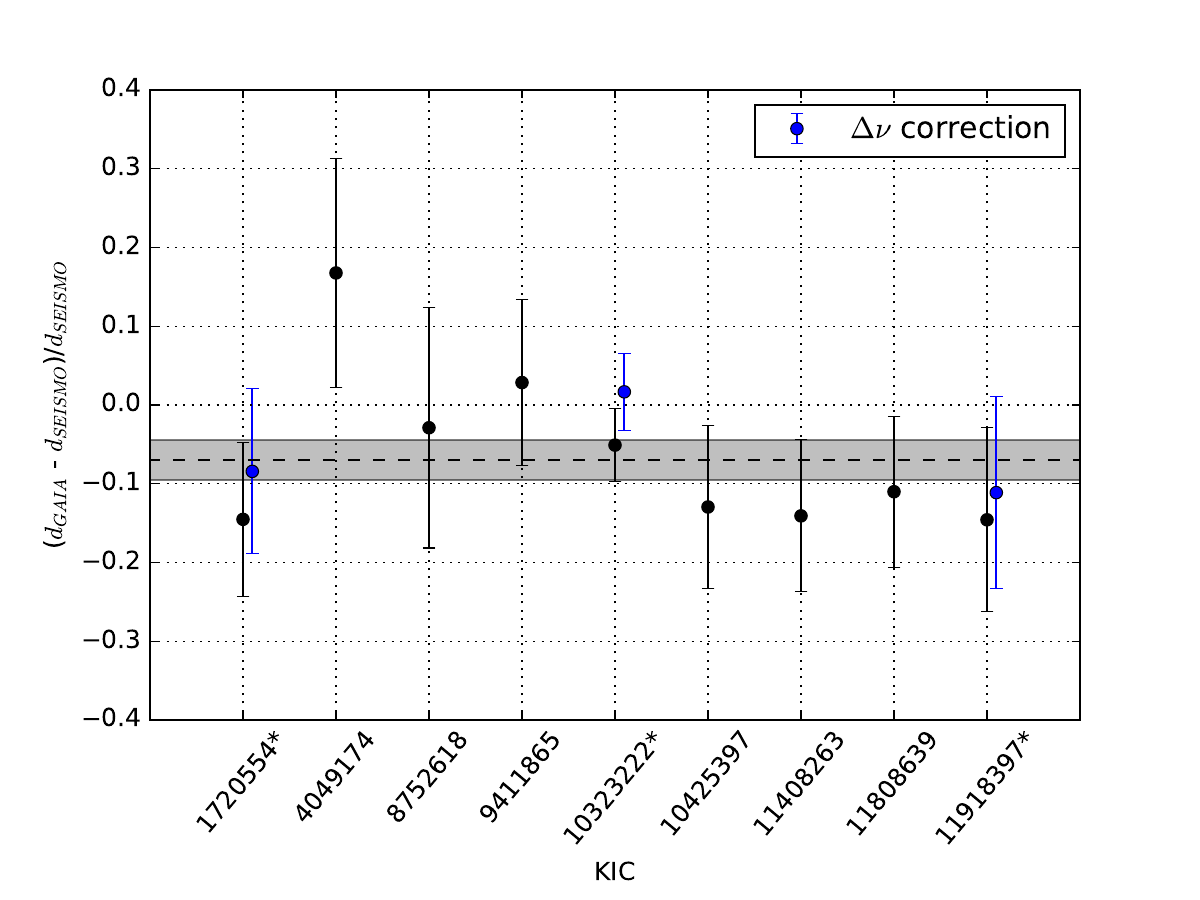} 
 \includegraphics[width=\columnwidth]{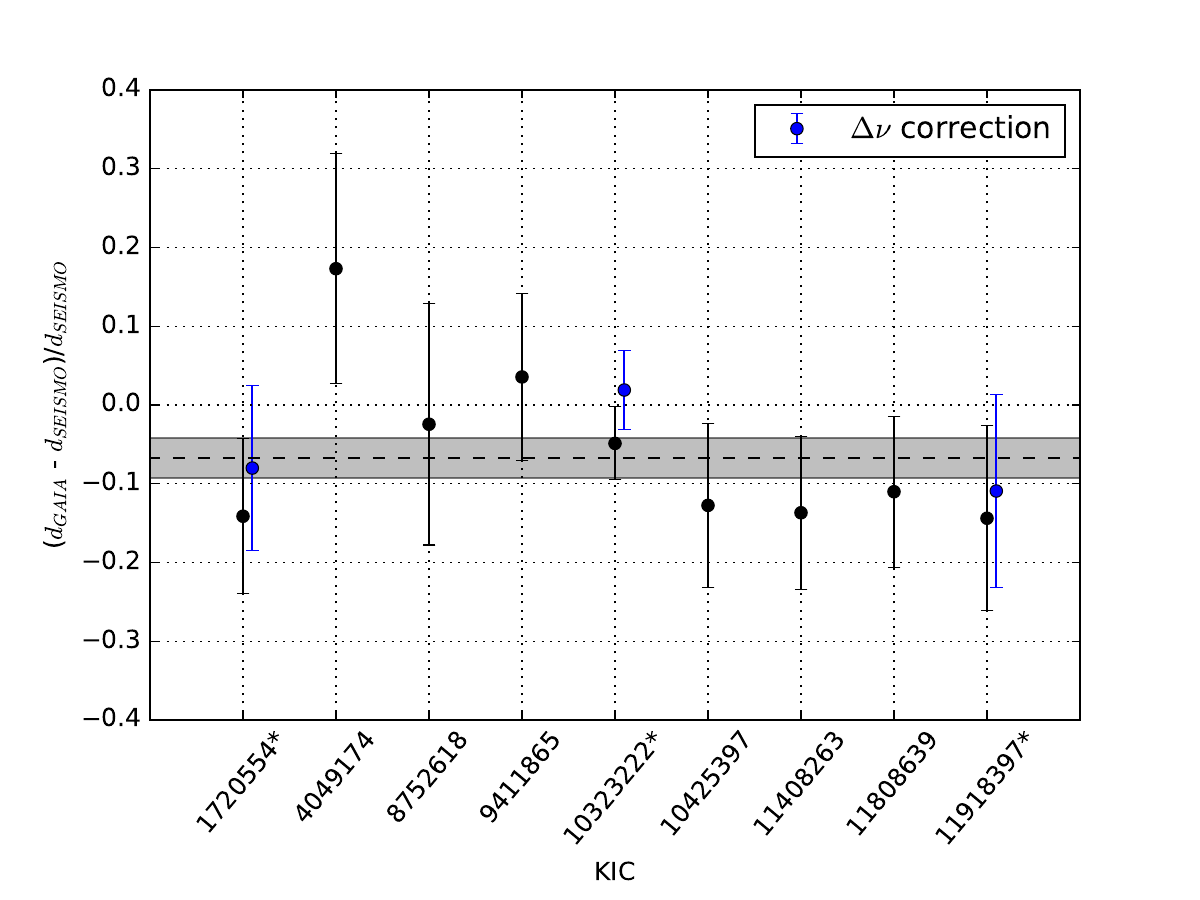}

 \caption{Left panel: Relative differences between distances derived from \textit{Gaia} DR3 and seismic distances (black dots) estimated using equation \ref{seismodist}. Effects of the applied correction to $\Delta\nu$ in the seismic distances are shown as blue dots. Seismic distances were calculated by using 2MASS J magnitudes, while the reddening effects were derived using dust maps from \citet{Green18}. The effects of extinction are accounted for in the calculations of seismic distances. The black dashed line is the weighted average difference of $-0.0700 \pm 0.0253$. Right panel: Same as left panel, but using dust maps from \citet{Drimmel03} instead. The black dashed line is the weighted average difference of $-0.0673 \pm 0.0256$.}
 
 \label{res_dist_Jmag}
\end{figure*}

\begin{figure*}

 \includegraphics[width=\columnwidth]{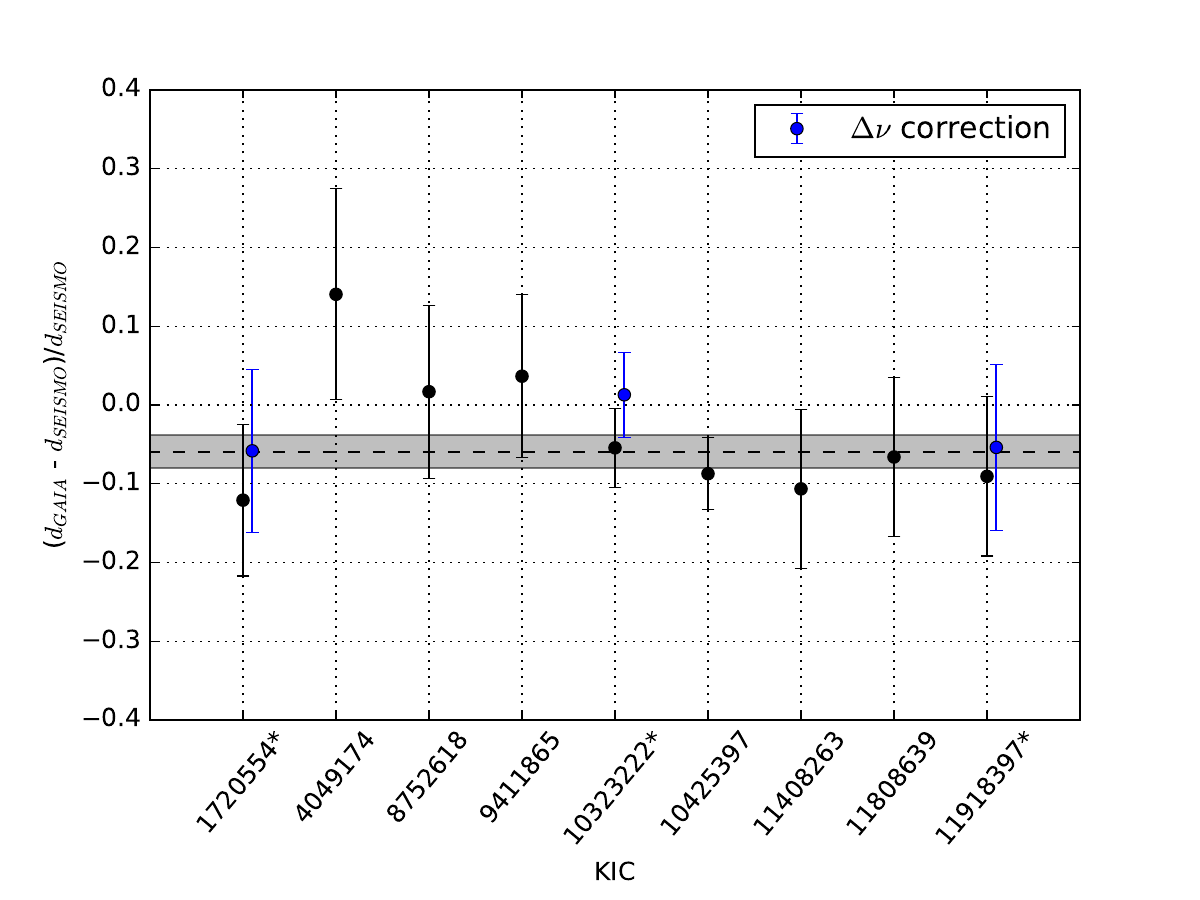} 
 \includegraphics[width=\columnwidth]{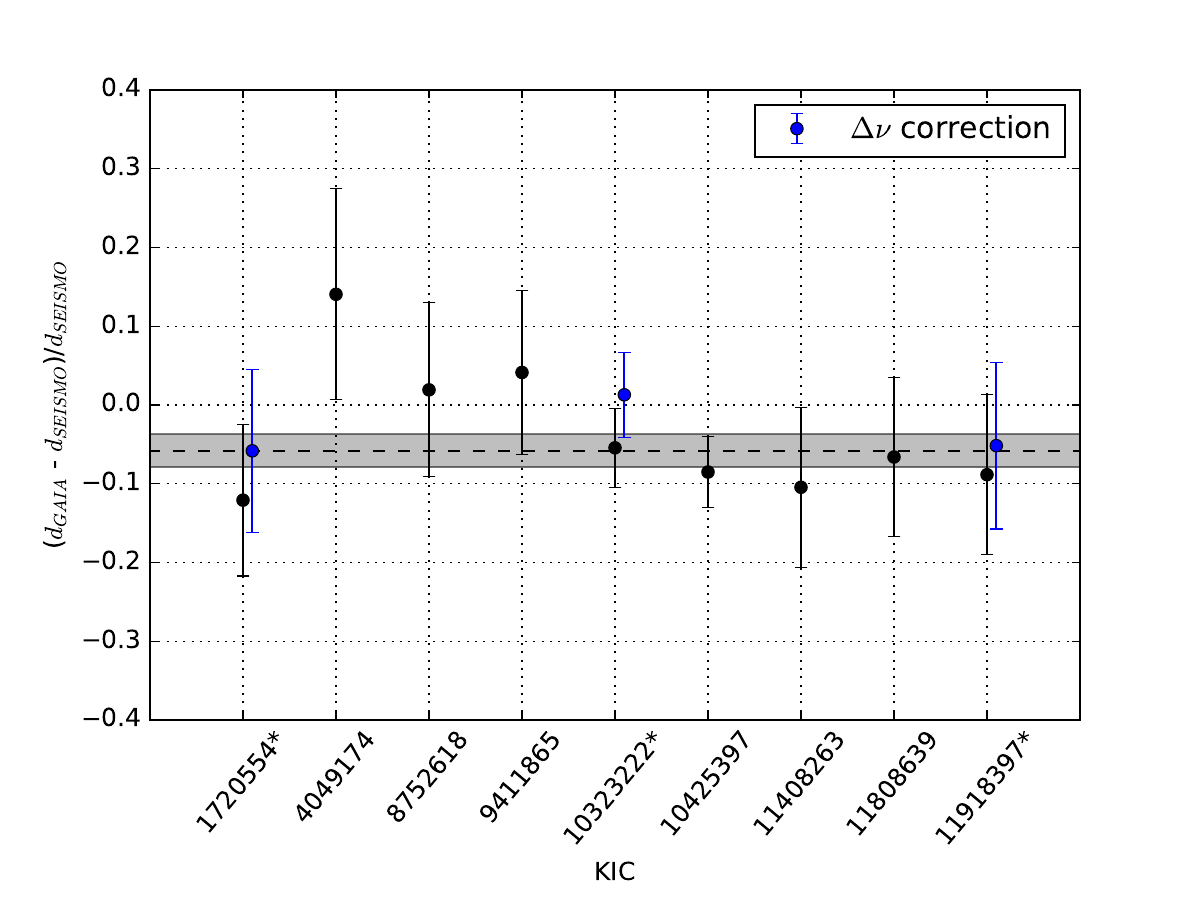}

 \caption{Same as Figure \ref{res_dist_Jmag}, but using magnitudes from 2MASS H band to calculate seismic distances. Left panel: Residual differences using dust maps from \citet{Green18}. The black dashed line is the weighted average difference of $-0.0592 \pm 0.0211$. Right panel: Residual differences using dust maps from \citet{Drimmel03}. The black dashed line is the weighted average difference of $-0.0582 \pm 0.0211$.}
 
 \label{res_dist_Hmag}
\end{figure*}

\begin{figure*}

 \includegraphics[width=\columnwidth]{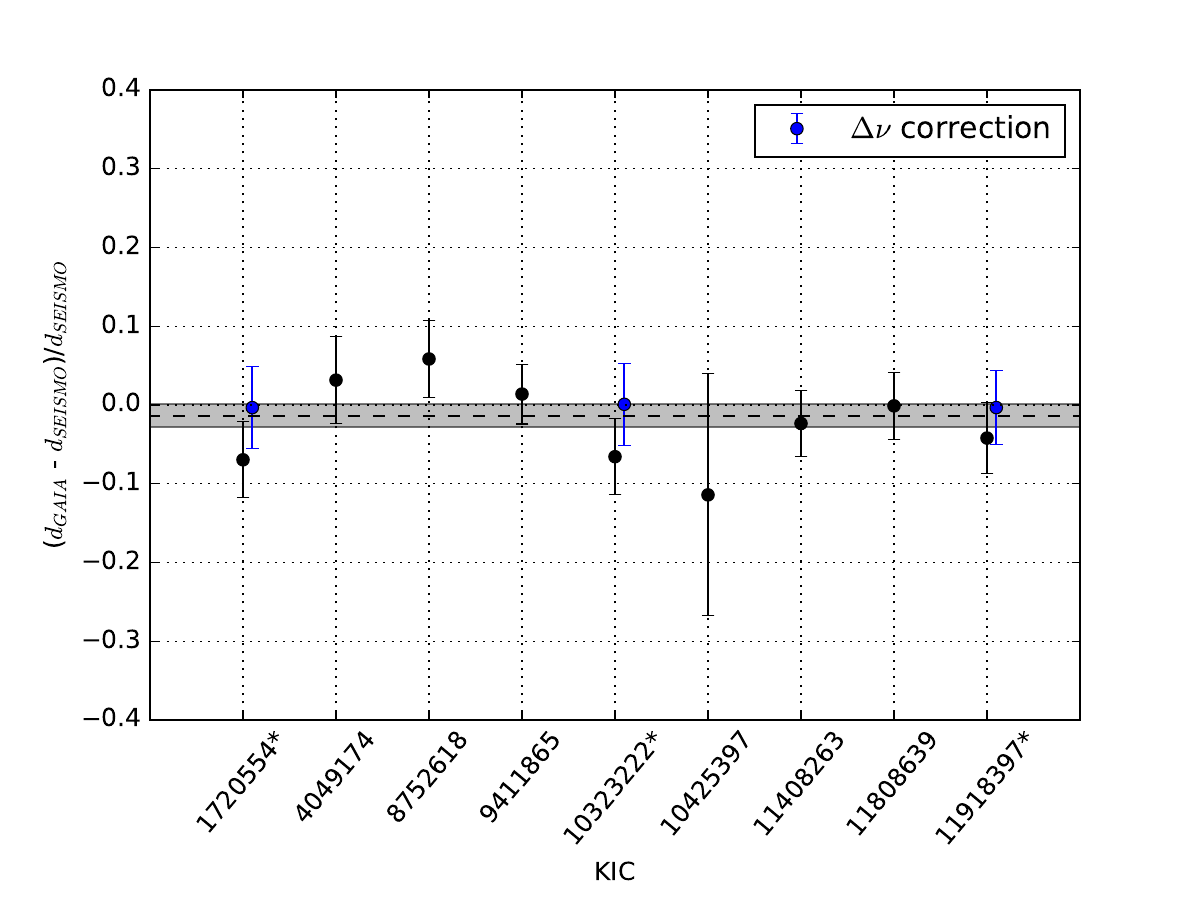} 
 \includegraphics[width=\columnwidth]{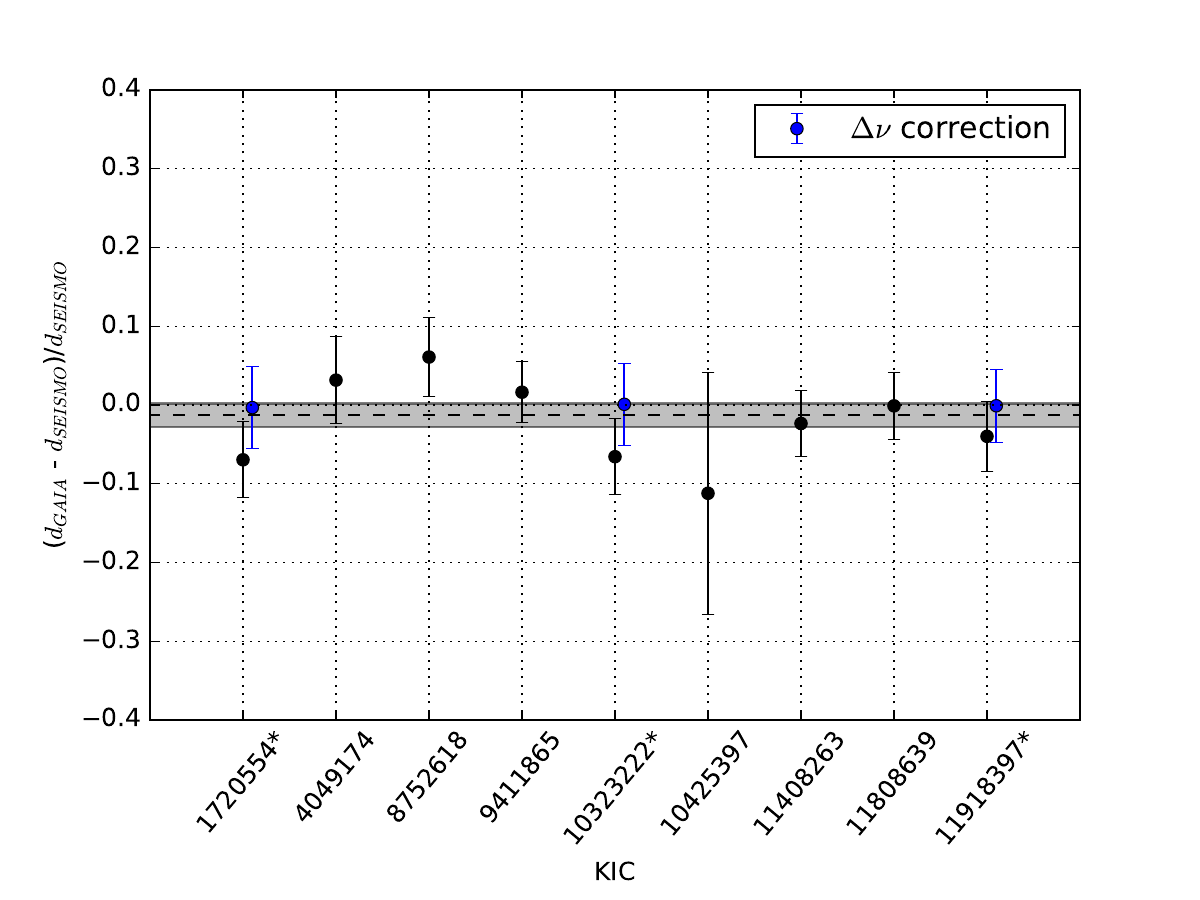}

 \caption{Same as Figure \ref{res_dist_Jmag}, but using magnitudes from 2MASS Ks band to calculate seismic distances. Left panel: Residual differences using dust maps from \citet{Green18}. The black dashed line is the weighted average difference of $-0.0134 \pm 0.0150$. Right panel: Residual differences using dust maps from \citet{Drimmel03}. The black dashed line is the weighted average difference of $-0.0131 \pm 0.0150$.}
 
 \label{res_dist_Kmag}
\end{figure*}

\begin{figure*}

 \includegraphics[width=\columnwidth]{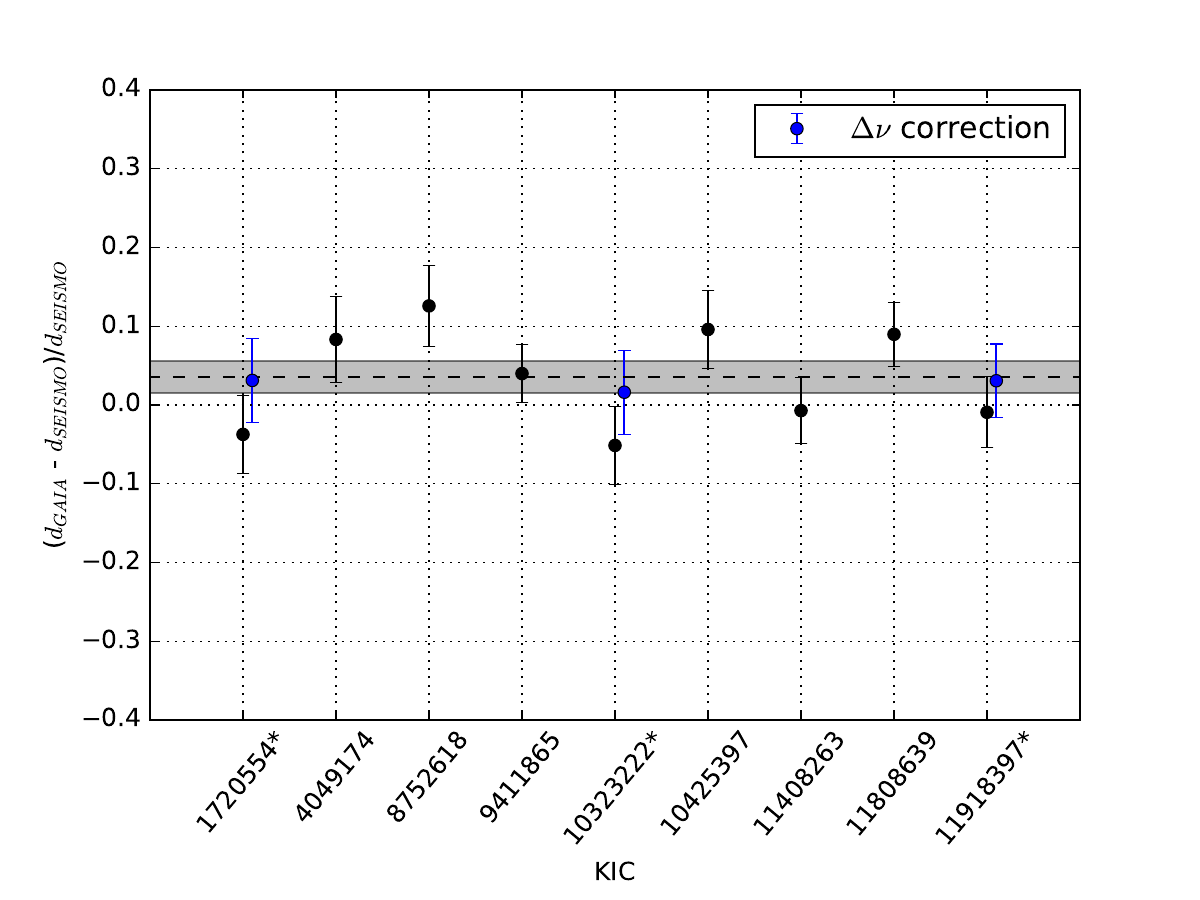} 
 \includegraphics[width=\columnwidth]{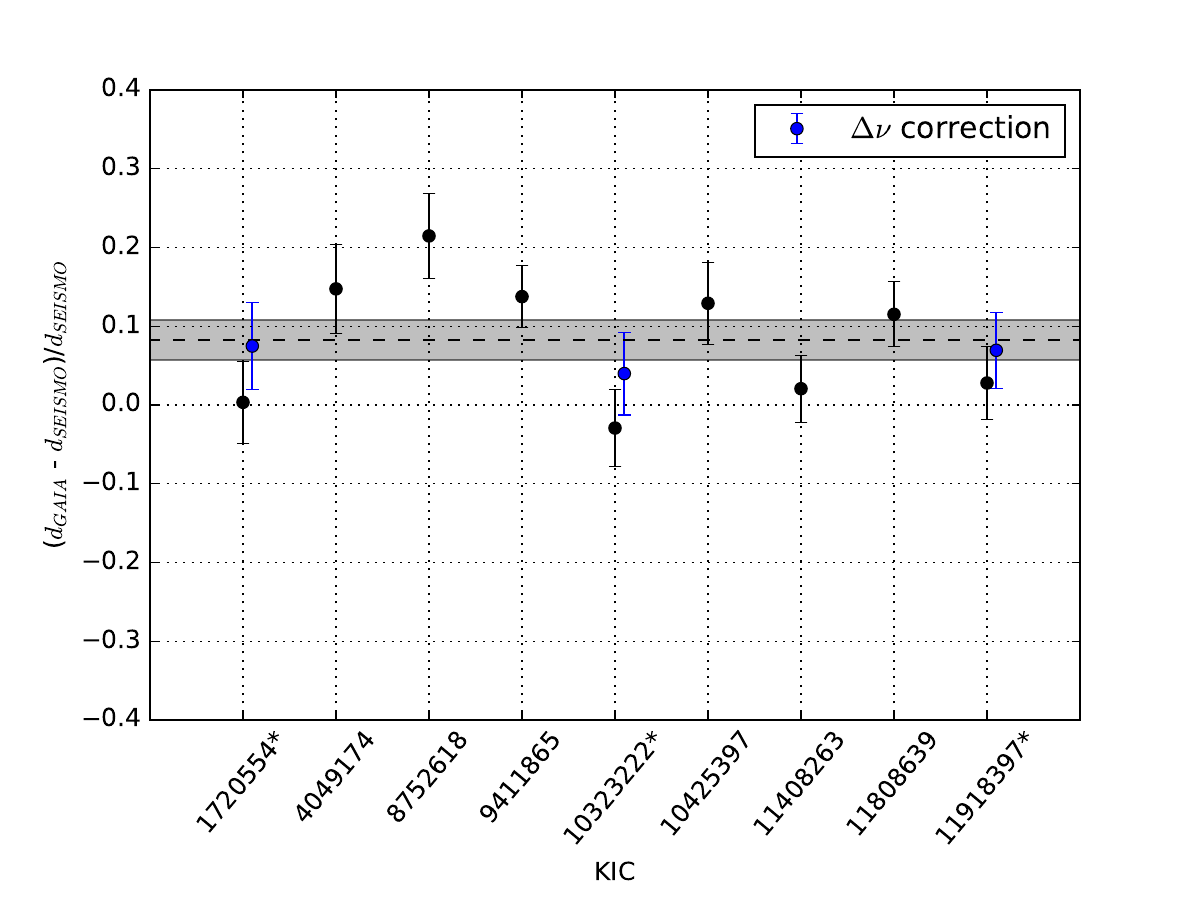}

 \caption{Same as Figure \ref{res_dist_Jmag}, but using V band magnitudes to calculate seismic distances. Left panel: Residual differences using dust maps from \citet{Green18}. The black dashed line is the weighted average difference of $0.0354 \pm 0.0202$. Right panel: Residual differences using dust maps from \citet{Drimmel03}. The black dashed line is the weighted average difference of $0.0828 \pm 0.0257$.}
 
 \label{res_dist_Vmag}
\end{figure*} 

\clearpage 

\end{document}